\documentclass[aps,prb,reprint,twocolumn]{revtex4-2}

\usepackage{graphicx}
\usepackage{bm}
\usepackage{hyperref}
\usepackage{amsmath}
\usepackage{amssymb}
\usepackage{mathtools}
\usepackage{siunitx}

\usepackage{comment}
\usepackage[table,dvipsnames,svgnames]{xcolor}
\usepackage{array}
\usepackage{multirow}
\usepackage{physics}
\usepackage[compat=1.0.0]{tikz-feynman}
\usepackage[caption=false]{subfig}
\usepackage{braket}

\usepackage[english]{babel}
\usepackage[autostyle, english = american]{csquotes}

\usepackage[scr=kp]{mathalpha}

\makeatletter
\def\maketitle{
\@author@finish
\title@column\titleblock@produce
\suppressfloats[t]}
\makeatother

\MakeOuterQuote{"}

\newcommand{\li}{\mathcal{C}}

\newcommand{\ee}{e}
\newcommand{\eqret}{\notag \\}  

\newcommand{\ii}{i}

\begin{document}
  \title{How pairing mechanism dictates topology in valley-polarized superconductors with Berry curvature}

\author{Julian May-Mann} 
\email{maymann@stanford.edu}
\author{Tobias Helbig}
\author{Trithep Devakul}
\email{tdevakul@stanford.edu}
\affiliation{Department of Physics, Stanford University, Stanford, CA 94305, USA}

\begin{abstract}
We investigate how the pairing mechanism influences topological superconductivity in valley-polarized systems with Berry curvature.
We demonstrate that short-range attractive interactions, such as those mediated by phonons, favor superconducting states where the Bogoliubov-de Gennes (BdG) Chern number has the same sign as the Berry curvature. 
In contrast, overscreened repulsive interactions, as in the Kohn-Luttinger mechanism, favor superconducting states where the BdG Chern number has the opposite sign as the Berry curvature.
We establish these trends in a fully controlled limit and apply them to a recently reported chiral superconductor in rhombohedral multilayer graphene.
Our theory provides a concrete experimental criterion for distinguishing between different pairing mechanisms in valley-polarized topological superconductors.

\end{abstract}
\maketitle

Understanding the fundamental pairing mechanism driving superconductivity in strongly correlated two-dimensional materials remains one of the most important questions in condensed matter physics.
In graphene-based superconductors (e.g., moiré twisted graphene multilayers~\cite{Balents2020Superconductivity, cao2018unconventional,yankowitz2019tuning, lu2019superconductors, arora2020superconductivity, park2021tunable, cao2021nematicity, hao2021electric, cao2021pauli, oh2021evidence, kim2022evidence, liu2022isospin, zhang2022promotion, park2022robust,su2023superconductivity}, Bernal bilayer graphene~\cite{zhou2022isospin,zhang2023enhanced,holleis2023nematicity, li2024tunable},  and rhombohedral stacked $N$-layer graphene (R$N$G)~\cite{zhou2021superconductivity, han2024signatures,choi2024electric}) proposed mechanisms have included conventional phonon-mediated pairing~\cite{peltonen2018mean, wu2019phonon, lian2019twisted, sarma2020electron, shavit2021theory, vinas2024phonon}, purely electronic mechanisms~\cite{sharma2020superconductivity, ghazaryan2021unconventional, you2022kohn, li2023charge, cea2022superconductivity}, isospin fluctuations~\cite{chatterjee2022inter,christos2023nodal,dong2023superconductivity,qin2023functional,dong2024superconductivity,fischer2024spin}, and more exotic possibilities~\cite{khalaf2021charged, cea2021coulomb,kwan2022skyrmions, kim2025topological}.
Understanding the underlying pairing mechanism in these systems could lead to insight into other correlated superconductors, such as cuprates, 
which have long defied explanation.

A recent experiment on R4G reported evidence of chiral superconductivity~\cite{han2024signatures}.
The normal state of this superconductor is a spin- and valley-polarized quarter-metal exhibiting an anomalous Hall effect, implying broken time-reversal symmetry (TRS) and high Berry curvature.  
A superconductor emerging from this normal state is a strong candidate for a topological superconductor --- hosting Majorana modes at vortex cores and chiral Majorana edge modes --- long sought after for potential applications in quantum information~\cite{qi2011topological, leijnse2012introduction, sato2017topological}. Such superconductors are indexed by an integer, $\li$, which is equal to the Chern number of the Bogoliubov-de-Gennes (BdG) Hamiltonian. 
Several recent theoretical studies suggest that the overscreening of repulsive interactions, i.e. the Kohn-Luttinger mechanism, is relevant for the chiral superconductor in R$4$G~\cite{geier2024chiral,shavit2024quantum, jahin2024enhanced, chou2024intravalley,yang2024topological,qin2024chiral,parra2025band}. 
However, experimentally distinguishing between possible mechanisms is inherently difficult and usually involves probing the superconductor's response to varying external parameters~\cite{saito2020independent, stepanov2020untying,liu2021tuning,barrier2024coulomb}.

A key aspect of this problem is the role of Berry curvature.
In R$N$G, the low-energy electronic bands feature a high valley-contrasting Berry curvature~\cite{koshino2009trigonal,zhang2010band}.  
In conventional superconductors, Cooper pairs form between electrons from opposite valleys, resulting in a cancellation of Berry phase.
In contrast, when the normal state is valley-polarized, pairing occurs within a single valley, thus making Berry phase effects essential.  
Given that Berry curvature, $\mathcal{B}(\bm{k})$, behaves like a magnetic field in momentum space~\cite{adams1959energy, nagaosa2010anomalous, price2014quantum}, one might expect it to act like an orbital Zeeman field, aligning the Cooper pair orbital angular momentum, $\ell$, with $\mathcal{B}$. 
Since $\li=\ell$ in a rotationally symmetric topological superconductor, this would suggest that the sign of $\li$ is dictated solely by $\mathcal{B}$. 
However, as we will show, this intuition fails in a striking way.

\begin{figure}[t!]
    \centering
    \includegraphics[width=\linewidth]{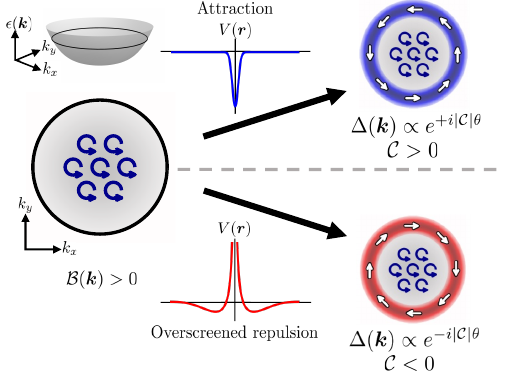}
    \caption{Summary of main result.
    Superconductivity in a band with positive Berry curvature $\mathcal{B}(\bm{k})$, illustrated by clockwise blue arrows. 
    Short-range attraction, such as that mediated by phonons, lead to $\li > 0$ superconductivity, aligned with $\mathcal{B}$.
    In contrast, overscreened repulsive interactions, as in the Kohn-Luttinger mechanism, lead to $\li < 0$ superconductivity anti-aligned with $\mathcal{B}$. White arrows show the winding of the gap function.}
    \label{fig:Schematic}
\end{figure}
In this work, we analyze how Berry curvature influences the topology of valley-polarized superconductors.  
We reveal that $\li$ is not simply dictated by $\mathcal{B}$ but also depends crucially on the nature of the pairing mechanism.  As summarized in Fig~\ref{fig:Schematic}, $\li$ aligns with $\mathcal{B}$ for pairing mediated by short-ranged attraction, such as in phonon-based theories. However, $\li$ and $\mathcal{B}$ tend to have \emph{opposite} signs when superconductivity arises from an overscreened Coulomb interaction, which is short-range repulsive but long-range attractive. 

To illustrate the counterintuitive nature of this result, consider a valley-polarized superconductor emerging from a single circular Fermi surface, as depicted in Fig~\ref{fig:Schematic}.  
In a smooth gauge, the phase winding of the momentum space order parameter $\Delta(\bm{k})$ around the Fermi surface is equal to $\li$.
Viewing $\Delta(\bm{k})$ as a superconducting ring in momentum space, with $\mathcal{B}$ acting as an effective magnetic field, a momentum space analog of the Little-Parks effect~\cite{PhysRevLett.9.9} suggests that $\li$ should scale roughly with the total Berry flux enclosed by the Fermi surface.  
While this intuition holds for short-range attraction, it fails dramatically for short-range repulsion --- instead producing an ``anti-Little-Parks'' effect where the phase winds \emph{against} the effective field!

This result has profound consequences.  Since $\li$ determines the number and chirality of Majorana edge modes, it controls the thermal Hall effect of the superconductor~\cite{read2000paired}.  
On the other hand, the anomalous Hall effect~\cite{nagaosa2010anomalous} of the normal state is determined by the total $\mathcal{B}$ enclosed by the Fermi surface.  
Our findings establish that the relative signs of the thermal and anomalous Hall effects, in the superconducting and normal states respectively, can serve as a key experimental test of the nature of the underlying pairing mechanism.

\begin{figure}[t!]
    \centering
    \includegraphics[width=\linewidth]{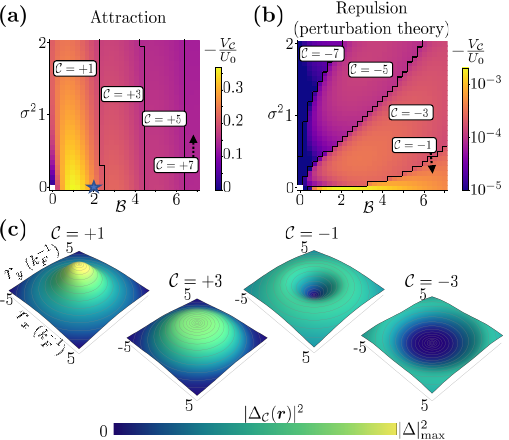}
    \caption{ Strength and BdG Chern number of the dominant pairing channel as a function of $\mathcal{B}$ and $\sigma^2$ for \textbf{(a)} short-range attraction and \textbf{(b)} weak short-range repulsion ($\frac{m}{\pi}U_0 = .1$) treated perturbatively. \textbf{(c)} $|\Delta_{\li}(\bm{r})|^2$, for $\li = \pm 1$ and $\pm 3$ pairing. Calculated using $\mathcal{B} = 2 $ and $\sigma^2 = 0$ (blue star in \textbf{a}) for pairing at the Fermi surface. $|\Delta|^2_{\text{max}} = |\Delta_{+1}(0)|^2$. }
    \label{fig:ICB_combined_plots} 
\end{figure}

\textit{Band projected Hamiltonian ---} 
We begin with an isolated band of spin- and valley-polarized electrons, corresponding to a quarter-metal phase.
The Hamiltonian is given by
\begin{align}\label{eq:FullHam}
    \hat{H} \!&= \!\sum_{\bm{k}} \epsilon(\bm{k}) \gamma^\dagger_{\bm{k}}\gamma^{\phantom{\dagger}}_{\bm{k}}  + \hat{H}_{\text{int}},\\ 
    \hat{H}_{\text{int}} &= \frac{1}{2A}\!\!\!\!\sum_{\bm{q},\bm{k}_1,\bm{k}_2} \!\!\! V^{\phantom{\dagger}}_{\bm{q}} \phantom{|} F^{\phantom{\dagger}}_{\bm{k}_1,\bm{k}_1\text{-}\bm{q}} \phantom{|}F^{\phantom{\dagger}}_{\bm{k}_2,\bm{k}_2+\bm{q}} \phantom{|}\gamma^\dagger_{\bm{k}_1}\gamma^{\phantom{\dagger}}_{\bm{k}_1\text{-}\bm{q}} \gamma^\dagger_{\bm{k}_2}\gamma^{\phantom{\dagger}}_{\bm{k}_2+\bm{q}}\notag
\end{align}
where $\gamma^\dagger$ are the band projected creation operators, $\epsilon(\bm{k})$ is the band dispersion, $\hat{H}_{\text{int}}$ describes a band-projected density-density interaction with potential $V_{\bm{q}}$, and $A$ is the system area. The Berry curvature $\mathcal{B}(\bm{k}) = i \epsilon^{ij} \partial_i \bra{u_{\bm{k}}}\partial_j \ket{u_{\bm{k}}}$ enters the Hamiltonian via the form factors, $F_{\bm{k_1},\bm{k}_2} = \braket{u_{\bm{k_1}}|u_{\bm{k}_2}}$, where $\ket{u_{\bm{k}}}$ is the periodic part of the Bloch wavefunction. 

\textit{Limit of uniform quantum geometry --- } Before moving on to more realistic and complex models, we will first analyze superconductivity in the limit of uniform quantum geometry. 
We consider a model with quadratic dispersion, $\epsilon(\bm{k}) = |\bm{k}|^2/2m$; short-ranged interactions, $V_{\bm{q}} = U_0$ (const.); and the form factors~\cite{tan2024parent}, 
\begin{equation}
    F^{\phantom{\dagger}}_{\bm{k_1},\bm{k}_2} =  e^{-\tfrac{\mathcal{B} + \sigma^2}{4}|\bm{k}_1-\bm{k}_2|^2 - i \tfrac{\mathcal{B}}{2} \left( \bm{k}_1\times\bm{k}_2 \right)}.
\label{eq:non-idealFF}\end{equation}
The parameter $\sigma^2$ allows for independent control of the  Berry curvature, $\mathcal{B} > 0$ and the quantum geometric tensor of the band~\cite{resta2011insulating}, $g_{ij} = \frac{1}{2}(\mathcal{B} + \sigma^2)\delta_{ij}$. When $\sigma^2 = 0$, the form factors match those of the lowest Landau level, and the band is said to have ``ideal'' quantum geometry~\cite{parameswaran2013fractional, roy2014band, jackson2015geometric}. 
We remark that this model is formally equivalent to one with ideal form factors and a Gaussian interaction $V_{\bm{q}} = U_0 \exp(- |\bm{q}|^2\sigma^2/2)$. 
Thus, $\sigma^2$ serves the dual purpose of tuning band geometry or controlling interaction range, depending on the interpretation.

This model is fully described by three dimensionless parameters, $\mathcal{B} k_F^2$, $\sigma^2 k_F^2$, and $\frac{m}{\pi}U_0$, where $k_F$ is the Fermi wavevector and $\frac{m}{\pi}$ is the density of states. 
These parameters control the Berry curvature, quantum metric, and interaction strength respectively. 
For convenience, we set $k_F = 1$ unless otherwise stated.

\textit{Attractive interactions ---}  
First, we consider superconductivity driven by an attractive interaction, $U_0<0$. 
To determine the BdG Chern number associated with the dominant superconducting instability, we project the pairing interaction to the Fermi surface. 
This amounts to setting $\bm{k}_1 = -\bm{k}_2 = \bm{k}$, $\bm{q} = \bm{k} - \bm{k}'$, and $|\bm{k}|=|\bm{k}'| = k_F$ in $\hat{H}_{\text{int}}$.
In a rotationally invariant system, if pairing is restricted to the Fermi surface, each superconducting instability is characterized by a unique winding number $\li\in\mathbb{Z}$.
The pairing interaction in the $\li$-channel is given by $V_{\li} = \int d\theta V(\theta) e^{-i \li \theta}$, where $V(\theta)$ is the band-projected interaction (including form factors) as a function of the angle $\theta$ between $\bm{k}$ and $\bm{k}'$. 
The dominant instability corresponds to the most negative $V_{\li}$ with odd $\li$, and leads to a superconductor with BdG Chern number $\li$. 
The transition temperature for this state is $T_c\propto \exp(- 1/\rho | V_{\li}| )$, where $\rho$ is the density of states at the Fermi surface.

For attractive interactions, $V_{\li}$ has the analytic form,
\begin{equation}\begin{split}
    & V_{\li} \! = U_0 e^{- \left [ \mathcal{B}+\sigma^2\right ]} \! \left ( \tfrac{2\mathcal{B}+\sigma^2}{\sigma^2}\right )^{\li/2} \! \! I_{\li}(\sqrt{\!2\mathcal{B}\sigma^2+\sigma^4})
\end{split}\end{equation}
where $I_{\li}$ is the modified Bessel function of order $\li$. In Fig.~\ref{fig:ICB_combined_plots}(a) we show $\li$ of the dominant instability, as well as the corresponding value of $V_{\li} $ as a function of $\mathcal{B}$ and $\sigma^2$. In agreement with the momentum space Little-Parks analogy, we find that $\li> 0$ for all instabilities, and the value is mostly determined by $\mathcal{B}$. 
The attraction is strongest for $\sigma^2 = 0$.

 \textit{Repulsive interactions ---} Next, we consider superconductors that emerge from the overscreening of a repulsive interaction. 
 To this end, we take $U_0 > 0$ in the bare interaction and calculate the overscreened effective interaction to order $U^2_0$ in perturbation theory~\cite{kohn1965new,raghu2010superconductivity,arovas2022hubbard}. 
 This calculation can be done analytically (see  SM~\cite{supp} for details).
 Furthermore, this perturbative approach is asymptotically \emph{exact} in the limit of small $U_0$~\cite{raghu2010superconductivity}.
 Using the overscreened interaction, we can then determine the dominant superconducting instability by following the same logic as before. The results are shown in Fig~\ref{fig:ICB_combined_plots}(b). 
 Strikingly, we find $\li <0$ superconductivity is favored for all parameters considered.

 Interestingly, Fig~\ref{fig:ICB_combined_plots}(b) shows that finite Berry curvature is essential for robust pairing.
 This can be understood as, in the $\mathcal{B} \rightarrow 0$ and $\sigma^2 \rightarrow 0$ limit, the Hamiltonian reduces to that of a conventional 2D electron gas (2DEG) with a short-range interaction. 
 It is well known that Kohn-Luttinger superconductivity does not occur at order $U^2_0$ in such a 2DEG, instead occurring at order $U^3_0$~\cite{chubukov1993kohn}. 
By contrast, when $\mathcal{B}>0$, robust attractive channels emerge already at order $U^2_0$, a direct consequence of nonzero Berry curvature. 
Moreover, for both attractive and repulsive interactions, pairing is strongest when $\sigma=0$, suggesting that ideal band geometry enhances superconductivity regardless of pairing mechanism.

\textit{Microscopic origin ---} 
To understand the physical origin of the trends we have observed so far, let us consider superconductivity from a real-space perspective. 
For a given interaction, the electrons that make up a Cooper pair arrange themselves to maximize attraction and minimize repulsion. 
Short-range attractive interactions naturally favor Cooper pairs in which electrons are close together.
In contrast, overscreened repulsive interactions can be understood as attractive at long distances, but repulsive at short distances (see Fig.~\ref{fig:Schematic}).
As a result, electrons in these Cooper pairs prefer to stay further apart to avoid the short-range repulsion. 

Now, consider the Cooper pairs associated with phase winding $\pm |\li|$. 
For time-reversal symmetric bands (i.e., those with no Berry curvature), Cooper pairs with $\pm|\li|$ have identical spatial profiles. 
However, as we shall show, when $\mathcal{B}>0$, $+|\li|$ Cooper pairs exhibit a shorter inter-particle distance than $-|\li|$ Cooper pairs. 
This naturally explains why attractive interactions favor $\li>0$ superconductivity, while overscreened repulsion favors $\li<0$.

The real-space distribution of the Cooper pair is given by
\begin{equation}
    |\Delta(\bm{r})|^2 = \frac{1}{A^2} \sum_{\bm{k},\bm{k}'} \Delta(\bm{k})\Delta^*(\bm{k}')(F^{\phantom{\dagger}}_{\bm{k},\bm{k}'})^2 e^{i (\bm{k}-\bm{k}')\cdot \bm{r}},
\end{equation} 
where $\bm{r}$ is the inter-particle distance.
A key quantity to consider is $|\Delta(\bm{r}=0)|^2$, which we refer to as the ``local occupancy''.
For bands with trivial form factors (e.g. bands made up of a single microscopic orbital), Pauli exclusion enforces a local occupancy of zero.
However, in bands with non-trivial form factors, the local occupancy can be non-zero.

As an illustration, consider an order parameter $\Delta_{\li}(\bm{k})\sim e^{i\li \theta}\delta(|\bm{k}|-k_F)$ that is nonzero only near the Fermi surface.
In the limit $\sigma^2=0$, $|\Delta_{\li}(\bm{r})|^2$ takes the form
\begin{equation}\begin{split}
    |\Delta_{\li}(\bm{r})|^2  \propto \sum_{n\geq 0} e^{-\mathcal{B}} \frac{(\mathcal{B})^n}{n!}J_{|\li-n|}\left (k_F |\bm{r}|\right)^2
\label{eq:Delta_Spatial}\end{split}\end{equation}
up to an $\li$-independent prefactor. In Fig~\ref{fig:ICB_combined_plots}(c) we plot $|\Delta_{\li}(\bm{r})|^2$ for several choices of $\li$. 
In this limit, the local occupancy vanishes for all $-|\li|$ states and is nonzero for all $+|\li|$ states. 
When $\sigma^2>0$, the local occupancy is finite for both $\pm|\li|$ but, for fixed $\li$, the $-|\li|$ local occupancy is always smaller than the $+|\li|$ by a factor $(\frac{\sigma^2}{2\mathcal{B}+\sigma^2})^\li < 1$.

This directly explains the previously observed trends. 
Since $+|\li|$ Cooper pairs have a higher local occupancy and are therefore more tightly bound, they are favored by short-range attractive interactions.
Conversely, $-|\li|$ Cooper pairs are less tightly bound, so are more favored by the overscreened repulsive interactions.
Thus, the local occupancy provides a direct link between normal state band properties and the preferred superconducting topology under different pairing mechanisms.

In fact, we can prove that the $\li=-1$ local occupancy is \textit{always} lower than the $\li= +1$ local occupancy for any rotation invariant system with positive Berry curvature, in the dilute limit $k_F\rightarrow 0$. 
One can expand the form factors in harmonics of $\theta$, 
    $F_{\bm{k},\bm{k}'} = \sum_{n} f_n e^{-i n \theta}$,
for $|\bm{k}|=|\bm{k}'|=k_F$,
where $f_n$ are real, non-negative, and scale as $k_F^{|n|}$~\cite{supp}. 
In the dilute limit, we can therefore restrict the sum to $-1 \leq n \leq 1$. 
In this case, the $\li= \pm 1$ local occupancy is $\propto f_{\pm 1}f_0$. 
Since the Berry curvature enclosed by the Fermi surface is given by $2\pi(f_1 - f_{-1})$, we conclude that the $\li= -1$ local occupancy is lower whenever the Berry curvature is positive.

\begin{figure}[t!]
    \centering
    \includegraphics[width=\linewidth]{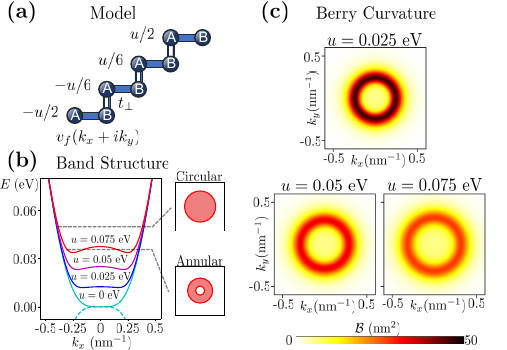}
    \caption{\textbf{(b)} Schematic of $h_{\text{R$4$G}}$.\textbf{(b)} Band structure of $h_{\text{R$4$G}}$ at various interlayer potentials $u$. Inserts show the low density annular Fermi surface and high density circular Fermi-surface for $u = .075$ eV.  \textbf{(c)} Berry curvature distribution of $h_{\text{R$4$G}}$.}
    \label{fig:R4G_BS_BC} 
\end{figure}

\textit{Rhombohedral graphene ---}
Having established the roles of both the Berry curvature and the pairing mechanism in determining the sign of $\li$ in the limit of uniform quantum geometry, we now demonstrate that these conclusions hold more generally in realistic settings.

We focus on spin- and valley-polarized R$4$G~\cite{han2024signatures} in the following analysis (results for R$N$G with $N=2,3,5,6$ exhibit qualitatively similar behavior~\cite{supp}).
We assume that flavor polarization occurs via the Stoner mechanism at a higher energy scale than superconductivity, justifying projection onto a single spin-valley component.
The choice of polarization into the $K$ or $K'$ valley occurs spontaneously, but can be trained by an external magnetic field~\cite{han2024signatures}.

We consider the $K$ valley of R$4$G modeled by the $8\times 8$ Bloch Hamiltonian,
\begin{equation}
    \begin{split}
    &h_{\text{R$4$G}}(\bm{k})_{2n-1,2n} = h_{\text{R$4$G}}^*(\bm{k})_{2n,2n-1} = v_f (k_x + i k_y)\\
    &h_{\text{R$4$G}}(\bm{k})_{2n,2n+1} = h_{\text{R$4$G}}^*(\bm{k})_{2n+1,2n} = -t_\perp\\
    &h_{\text{R$4$G}}(\bm{k})_{2n-1,2n-1} = h_{\text{R$4$G}}(\bm{k})_{2n,2n} = u_n
\label{eq:NlayerRhombo}\end{split}
\end{equation}
for $1 \leq n \leq 4$, where $v_f = 10^6$ m/s is the Fermi velocity, and $t_\perp = -\SI{.38}{eV}$~\cite{han2024signatures}, depicted in Fig~\ref{fig:R4G_BS_BC}(a).
The layer potential is given by $u_n=(-\frac{u}{2},-\frac{u}{6},\frac{u}{6},\frac{u}{2})$, where $u$ is a tuning parameter modeling the effect of a displacement field.  
We neglect longer-range hoppings, which induce trigonal warping effects, and as a result, $h_{\text{R$4$G}}$ retains continuous rotational symmetry.

We consider the superconductor emerging from the lowest electron band of R$4$G~\cite{han2024signatures}.
The complete Hamiltonian is Eq~\ref{eq:FullHam} with $\epsilon(\bm{k})$ and $F_{\bm{k}_1,\bm{k}_2}$ obtained by solving $h_{\text{R$4$G}}$.
This band exhibits several notable features. 
The dispersion, shown in Fig~\ref{fig:R4G_BS_BC}(b), leads to a Fermi surface that is annular at low densities but becomes circular at higher densities. 
The Berry curvature $\mathcal{B}(\bm{k})$ is positive and highly non-uniform,  forming a ring in momentum space, as shown in Fig~\ref{fig:R4G_BS_BC}(c).
For an annular Fermi surface, all superconducting instabilities result in BdG Chern number $\li=0$ regardless of the winding number.
We therefore use $\ell$ to refer to the winding number in this section, with the implicit understanding that $\ell=\li$ when the Fermi surface is circular.

\begin{figure}[b!]
    \centering
    \includegraphics[width=\linewidth]{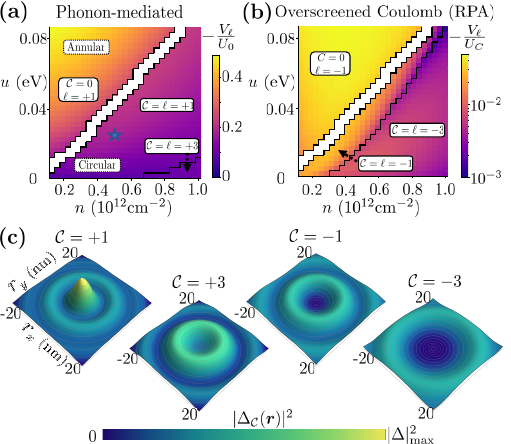}
    \caption{Strength of the dominant pairing channel as a function of electron density $n$ and $u$ for \textbf{(a)} phonon-mediated attraction and \textbf{(b)} an overscreened Coulomb interaction (calculated using RPA) for $d =20$nm and $\varepsilon = 4$, where $U_{C} = 1/\varepsilon d$. Regions with annular and circular Fermi-surfaces are separated.  
    \textbf{(c)} $|\Delta_{\li}(\bm{r})|^2$ for $\li = \pm 1$ and $\pm 3$ pairing, calculated for $u = \SI{.025}{eV}$ and $n = \SI{0.5e12}{cm^{-2}}$ (blue star in \textbf{a}). $|\Delta|^2_{\text{max}} = |\Delta_{+1}(0)|^2$}
    \label{fig:R4G_PD_WF} 
\end{figure}

 We consider two possible pairing mechanisms: phonon-mediated pairing and an overscreened Coulomb interaction. 
In the phonon scenario, the E$_2$ optical mode~\cite{basko2008interplay, wu2018theory} likely plays an important role, as it leads to intravalley scattering. 
Acoustic modes have also been argued to be important in related graphene superconductors~\cite{chou2021acoustic, chou2022acoustic, chou2022acoustic2}. 
Here, we neglect the detailed structure of the phonon modes and retardation effects, modeling the phonons as a short-range attractive interaction, $V_{\bm{q}} = U_0<0$.
For the overscreened Coulomb interaction, we adoped a bare repulsive potential based on a realistic dual-gated geometry,
\begin{equation}
    V_{\bm{q}} = 2\pi\tanh(d |\bm{q}|/2)/(\varepsilon|\bm{q}|),
\end{equation}
where $\varepsilon = 4$ and $d = 20$nm. 
Since $V_{\bm{q}}$ is not a weak perturbation, we incorporate overscreening effects within the random phase approximation (RPA).
We then analyze the leading superconducting instability using the same approach as before (see SM~\cite{supp} for details).

In Fig.~\ref{fig:R4G_PD_WF}(a,b) we plot the dominant superconducting instability as a function of $u$ and electron density $n$.  
Across all parameter values considered, phonon-mediated attraction favors $\ell>0$, while overscreened Coulomb interactions favor $\ell<0$,
consistent with our conclusions in the uniform quantum geometry limit.
This trend can be again directly attributed to differences in the real-space Cooper pair distribution, shown in Fig~\ref{fig:R4G_PD_WF}(c).

Several effects have been neglected in this analysis.
Including longer-range hopping terms introduces trigonal warping, reducing $SO(2)$ rotation symmetry to $C_3$~\cite{koshino2009trigonal}. 
Since this lifts the $\epsilon(\bm{k})=\epsilon(-\bm{k})$ degeneracy, pairing requires a critical attractive interaction strength.
Below $T_c$, resolving the competition between different states also necessitates solving the full BdG self-consistency equations.
Additionally, Hartree-Fock effects should renormalize the effective band structure.
Recent works have examined superconductivity in R$4$G, mediated by overscreened Coulomb interactions, accounting for some of these effects~\cite{yang2024topological,chou2024intravalley,qin2024chiral,parra2025band}.
Where reported~\cite{yang2024topological,qin2024chiral,parra2025band}, $\li$ is opposite in sign to $\mathcal{B}(\bm{k})$, in agreement with our analysis.
This indicates that, despite several simplifications, our theory correctly captures a general property of valley-polarized superconductors with Berry curvature.

\emph{Discussion ---}
We have demonstrated that the relative signs of $\li$ and $\mathcal{B}$ depend on the underlying pairing mechanism.  
This relationship holds in explicit examples and is supported by general arguments based on the spatial structure of Cooper pairs.
However, this result is not rigorous, and fine-tuned counterexamples exist~\cite{supp}.
Nevertheless, this trend persists across the entire range of realistic scenarios we have explored.

An immediate consequence of our theory is that the relative signs of the anomalous and the thermal Hall effects, in the normal and superconducting states respectively, can serve as a direct indicator of the microscopic pairing mechanism.
While the precise magnitudes of these effects depend on details, their relative signs are robust.
To ensure consistency, these effects must be measured in states with the same valley polarization, which can be stabilized by a small positive magnetic field. 
In R$4$G, at least one of the superconductors observed in Ref.~\cite{han2024signatures} is believed to originate from a single Fermi surface.
If pairing is mediated by an unconventional Kohn-Luttinger mechanism, as suggested by recent theories~\cite{yang2024topological,shavit2024quantum, jahin2024enhanced, chou2024intravalley,qin2024chiral,geier2024chiral,parra2025band}, our theory predicts that the anomalous and thermal Hall effects should exhibit opposite signs.

\emph{Conclusions ---} Our results elucidate how both Berry curvature and pairing mechanism play a decisive role in shaping superconducting topology.
In particular, the spatial structure of Cooper pairs provides a simple yet powerful criterion for determining which topological superconducting states are energetically favored under different pairing mechanisms.
These insights provide a tangible experimental criterion for distinguishing between different pairing mechanisms, offering a guiding principle for understanding and characterizing novel topological superconductors.

\textit{Acknowledgements} --- We thank Yiming Wu, Sri Raghu, Gal Shavit, Mike Zaletel, Long Ju, Sid Parameswaran, Tobias Holder, and Emily Zhang for useful conversations. JMM and TD were supported by a startup fund at Stanford University. TH was supported by the Deutsche Forschungsgemeinschaft (DFG, German Research Foundation) under Project No. 537357978 and, in part, by the US Department of Energy, Office of Basic Energy Sciences, Division of Materials Sciences and Engineering, under Contract No. DE-AC02-76SF00515.

\bibliography{KL_with_BC}

\begin{thebibliography}{79}%
\makeatletter
\providecommand \@ifxundefined [1]{%
 \@ifx{#1\undefined}
}%
\providecommand \@ifnum [1]{%
 \ifnum #1\expandafter \@firstoftwo
 \else \expandafter \@secondoftwo
 \fi
}%
\providecommand \@ifx [1]{%
 \ifx #1\expandafter \@firstoftwo
 \else \expandafter \@secondoftwo
 \fi
}%
\providecommand \natexlab [1]{#1}%
\providecommand \enquote  [1]{``#1''}%
\providecommand \bibnamefont  [1]{#1}%
\providecommand \bibfnamefont [1]{#1}%
\providecommand \citenamefont [1]{#1}%
\providecommand \href@noop [0]{\@secondoftwo}%
\providecommand \href [0]{\begingroup \@sanitize@url \@href}%
\providecommand \@href[1]{\@@startlink{#1}\@@href}%
\providecommand \@@href[1]{\endgroup#1\@@endlink}%
\providecommand \@sanitize@url [0]{\catcode `\\12\catcode `\$12\catcode
  `\&12\catcode `\#12\catcode `\^12\catcode `\_12\catcode `\%12\relax}%
\providecommand \@@startlink[1]{}%
\providecommand \@@endlink[0]{}%
\providecommand \url  [0]{\begingroup\@sanitize@url \@url }%
\providecommand \@url [1]{\endgroup\@href {#1}{\urlprefix }}%
\providecommand \urlprefix  [0]{URL }%
\providecommand \Eprint [0]{\href }%
\providecommand \doibase [0]{http://dx.doi.org/}%
\providecommand \selectlanguage [0]{\@gobble}%
\providecommand \bibinfo  [0]{\@secondoftwo}%
\providecommand \bibfield  [0]{\@secondoftwo}%
\providecommand \translation [1]{[#1]}%
\providecommand \BibitemOpen [0]{}%
\providecommand \bibitemStop [0]{}%
\providecommand \bibitemNoStop [0]{.\EOS\space}%
\providecommand \EOS [0]{\spacefactor3000\relax}%
\providecommand \BibitemShut  [1]{\csname bibitem#1\endcsname}%
\let\auto@bib@innerbib\@empty
\bibitem [{\citenamefont {Balents}\ \emph {et~al.}(2020)\citenamefont
  {Balents}, \citenamefont {Dean}, \citenamefont {Efetov} \emph
  {et~al.}}]{Balents2020Superconductivity}%
  \BibitemOpen
  \bibfield  {author} {\bibinfo {author} {\bibfnamefont {L.}~\bibnamefont
  {Balents}}, \bibinfo {author} {\bibfnamefont {C.~R.}\ \bibnamefont {Dean}},
  \bibinfo {author} {\bibfnamefont {D.~K.}\ \bibnamefont {Efetov}},  \emph
  {et~al.},\ }\href {\doibase 10.1038/s41567-020-0906-9} {\bibfield  {journal}
  {\bibinfo  {journal} {Nature Physics}\ }\textbf {\bibinfo {volume} {16}},\
  \bibinfo {pages} {725} (\bibinfo {year} {2020})}\BibitemShut {NoStop}%
\bibitem [{\citenamefont {Cao}\ \emph {et~al.}(2018)\citenamefont {Cao},
  \citenamefont {Fatemi}, \citenamefont {Fang}, \citenamefont {Watanabe},
  \citenamefont {Taniguchi}, \citenamefont {Kaxiras},\ and\ \citenamefont
  {Jarillo-Herrero}}]{cao2018unconventional}%
  \BibitemOpen
  \bibfield  {author} {\bibinfo {author} {\bibfnamefont {Y.}~\bibnamefont
  {Cao}}, \bibinfo {author} {\bibfnamefont {V.}~\bibnamefont {Fatemi}},
  \bibinfo {author} {\bibfnamefont {S.}~\bibnamefont {Fang}}, \bibinfo {author}
  {\bibfnamefont {K.}~\bibnamefont {Watanabe}}, \bibinfo {author}
  {\bibfnamefont {T.}~\bibnamefont {Taniguchi}}, \bibinfo {author}
  {\bibfnamefont {E.}~\bibnamefont {Kaxiras}}, \ and\ \bibinfo {author}
  {\bibfnamefont {P.}~\bibnamefont {Jarillo-Herrero}},\ }\href@noop {}
  {\bibfield  {journal} {\bibinfo  {journal} {Nature}\ }\textbf {\bibinfo
  {volume} {556}},\ \bibinfo {pages} {43} (\bibinfo {year} {2018})}\BibitemShut
  {NoStop}%
\bibitem [{\citenamefont {Yankowitz}\ \emph {et~al.}(2019)\citenamefont
  {Yankowitz}, \citenamefont {Chen}, \citenamefont {Polshyn}, \citenamefont
  {Zhang}, \citenamefont {Watanabe}, \citenamefont {Taniguchi}, \citenamefont
  {Graf}, \citenamefont {Young},\ and\ \citenamefont
  {Dean}}]{yankowitz2019tuning}%
  \BibitemOpen
  \bibfield  {author} {\bibinfo {author} {\bibfnamefont {M.}~\bibnamefont
  {Yankowitz}}, \bibinfo {author} {\bibfnamefont {S.}~\bibnamefont {Chen}},
  \bibinfo {author} {\bibfnamefont {H.}~\bibnamefont {Polshyn}}, \bibinfo
  {author} {\bibfnamefont {Y.}~\bibnamefont {Zhang}}, \bibinfo {author}
  {\bibfnamefont {K.}~\bibnamefont {Watanabe}}, \bibinfo {author}
  {\bibfnamefont {T.}~\bibnamefont {Taniguchi}}, \bibinfo {author}
  {\bibfnamefont {D.}~\bibnamefont {Graf}}, \bibinfo {author} {\bibfnamefont
  {A.~F.}\ \bibnamefont {Young}}, \ and\ \bibinfo {author} {\bibfnamefont
  {C.~R.}\ \bibnamefont {Dean}},\ }\href@noop {} {\bibfield  {journal}
  {\bibinfo  {journal} {Science}\ }\textbf {\bibinfo {volume} {363}},\ \bibinfo
  {pages} {1059} (\bibinfo {year} {2019})}\BibitemShut {NoStop}%
\bibitem [{\citenamefont {Lu}\ \emph {et~al.}(2019)\citenamefont {Lu},
  \citenamefont {Stepanov}, \citenamefont {Yang}, \citenamefont {Xie},
  \citenamefont {Aamir}, \citenamefont {Das}, \citenamefont {Urgell},
  \citenamefont {Watanabe}, \citenamefont {Taniguchi}, \citenamefont {Zhang}
  \emph {et~al.}}]{lu2019superconductors}%
  \BibitemOpen
  \bibfield  {author} {\bibinfo {author} {\bibfnamefont {X.}~\bibnamefont
  {Lu}}, \bibinfo {author} {\bibfnamefont {P.}~\bibnamefont {Stepanov}},
  \bibinfo {author} {\bibfnamefont {W.}~\bibnamefont {Yang}}, \bibinfo {author}
  {\bibfnamefont {M.}~\bibnamefont {Xie}}, \bibinfo {author} {\bibfnamefont
  {M.~A.}\ \bibnamefont {Aamir}}, \bibinfo {author} {\bibfnamefont
  {I.}~\bibnamefont {Das}}, \bibinfo {author} {\bibfnamefont {C.}~\bibnamefont
  {Urgell}}, \bibinfo {author} {\bibfnamefont {K.}~\bibnamefont {Watanabe}},
  \bibinfo {author} {\bibfnamefont {T.}~\bibnamefont {Taniguchi}}, \bibinfo
  {author} {\bibfnamefont {G.}~\bibnamefont {Zhang}},  \emph {et~al.},\
  }\href@noop {} {\bibfield  {journal} {\bibinfo  {journal} {Nature}\ }\textbf
  {\bibinfo {volume} {574}},\ \bibinfo {pages} {653} (\bibinfo {year}
  {2019})}\BibitemShut {NoStop}%
\bibitem [{\citenamefont {Arora}\ \emph {et~al.}(2020)\citenamefont {Arora},
  \citenamefont {Polski}, \citenamefont {Zhang}, \citenamefont {Thomson},
  \citenamefont {Choi}, \citenamefont {Kim}, \citenamefont {Lin}, \citenamefont
  {Wilson}, \citenamefont {Xu}, \citenamefont {Chu} \emph
  {et~al.}}]{arora2020superconductivity}%
  \BibitemOpen
  \bibfield  {author} {\bibinfo {author} {\bibfnamefont {H.~S.}\ \bibnamefont
  {Arora}}, \bibinfo {author} {\bibfnamefont {R.}~\bibnamefont {Polski}},
  \bibinfo {author} {\bibfnamefont {Y.}~\bibnamefont {Zhang}}, \bibinfo
  {author} {\bibfnamefont {A.}~\bibnamefont {Thomson}}, \bibinfo {author}
  {\bibfnamefont {Y.}~\bibnamefont {Choi}}, \bibinfo {author} {\bibfnamefont
  {H.}~\bibnamefont {Kim}}, \bibinfo {author} {\bibfnamefont {Z.}~\bibnamefont
  {Lin}}, \bibinfo {author} {\bibfnamefont {I.~Z.}\ \bibnamefont {Wilson}},
  \bibinfo {author} {\bibfnamefont {X.}~\bibnamefont {Xu}}, \bibinfo {author}
  {\bibfnamefont {J.-H.}\ \bibnamefont {Chu}},  \emph {et~al.},\ }\href@noop {}
  {\bibfield  {journal} {\bibinfo  {journal} {Nature}\ }\textbf {\bibinfo
  {volume} {583}},\ \bibinfo {pages} {379} (\bibinfo {year}
  {2020})}\BibitemShut {NoStop}%
\bibitem [{\citenamefont {Park}\ \emph {et~al.}(2021)\citenamefont {Park},
  \citenamefont {Cao}, \citenamefont {Watanabe}, \citenamefont {Taniguchi},\
  and\ \citenamefont {Jarillo-Herrero}}]{park2021tunable}%
  \BibitemOpen
  \bibfield  {author} {\bibinfo {author} {\bibfnamefont {J.~M.}\ \bibnamefont
  {Park}}, \bibinfo {author} {\bibfnamefont {Y.}~\bibnamefont {Cao}}, \bibinfo
  {author} {\bibfnamefont {K.}~\bibnamefont {Watanabe}}, \bibinfo {author}
  {\bibfnamefont {T.}~\bibnamefont {Taniguchi}}, \ and\ \bibinfo {author}
  {\bibfnamefont {P.}~\bibnamefont {Jarillo-Herrero}},\ }\href@noop {}
  {\bibfield  {journal} {\bibinfo  {journal} {Nature}\ }\textbf {\bibinfo
  {volume} {590}},\ \bibinfo {pages} {249} (\bibinfo {year}
  {2021})}\BibitemShut {NoStop}%
\bibitem [{\citenamefont {Cao}\ \emph {et~al.}(2021{\natexlab{a}})\citenamefont
  {Cao}, \citenamefont {Rodan-Legrain}, \citenamefont {Park}, \citenamefont
  {Yuan}, \citenamefont {Watanabe}, \citenamefont {Taniguchi}, \citenamefont
  {Fernandes}, \citenamefont {Fu},\ and\ \citenamefont
  {Jarillo-Herrero}}]{cao2021nematicity}%
  \BibitemOpen
  \bibfield  {author} {\bibinfo {author} {\bibfnamefont {Y.}~\bibnamefont
  {Cao}}, \bibinfo {author} {\bibfnamefont {D.}~\bibnamefont {Rodan-Legrain}},
  \bibinfo {author} {\bibfnamefont {J.~M.}\ \bibnamefont {Park}}, \bibinfo
  {author} {\bibfnamefont {N.~F.}\ \bibnamefont {Yuan}}, \bibinfo {author}
  {\bibfnamefont {K.}~\bibnamefont {Watanabe}}, \bibinfo {author}
  {\bibfnamefont {T.}~\bibnamefont {Taniguchi}}, \bibinfo {author}
  {\bibfnamefont {R.~M.}\ \bibnamefont {Fernandes}}, \bibinfo {author}
  {\bibfnamefont {L.}~\bibnamefont {Fu}}, \ and\ \bibinfo {author}
  {\bibfnamefont {P.}~\bibnamefont {Jarillo-Herrero}},\ }\href@noop {}
  {\bibfield  {journal} {\bibinfo  {journal} {science}\ }\textbf {\bibinfo
  {volume} {372}},\ \bibinfo {pages} {264} (\bibinfo {year}
  {2021}{\natexlab{a}})}\BibitemShut {NoStop}%
\bibitem [{\citenamefont {Hao}\ \emph {et~al.}(2021)\citenamefont {Hao},
  \citenamefont {Zimmerman}, \citenamefont {Ledwith}, \citenamefont {Khalaf},
  \citenamefont {Najafabadi}, \citenamefont {Watanabe}, \citenamefont
  {Taniguchi}, \citenamefont {Vishwanath},\ and\ \citenamefont
  {Kim}}]{hao2021electric}%
  \BibitemOpen
  \bibfield  {author} {\bibinfo {author} {\bibfnamefont {Z.}~\bibnamefont
  {Hao}}, \bibinfo {author} {\bibfnamefont {A.}~\bibnamefont {Zimmerman}},
  \bibinfo {author} {\bibfnamefont {P.}~\bibnamefont {Ledwith}}, \bibinfo
  {author} {\bibfnamefont {E.}~\bibnamefont {Khalaf}}, \bibinfo {author}
  {\bibfnamefont {D.~H.}\ \bibnamefont {Najafabadi}}, \bibinfo {author}
  {\bibfnamefont {K.}~\bibnamefont {Watanabe}}, \bibinfo {author}
  {\bibfnamefont {T.}~\bibnamefont {Taniguchi}}, \bibinfo {author}
  {\bibfnamefont {A.}~\bibnamefont {Vishwanath}}, \ and\ \bibinfo {author}
  {\bibfnamefont {P.}~\bibnamefont {Kim}},\ }\href@noop {} {\bibfield
  {journal} {\bibinfo  {journal} {Science}\ }\textbf {\bibinfo {volume}
  {371}},\ \bibinfo {pages} {1133} (\bibinfo {year} {2021})}\BibitemShut
  {NoStop}%
\bibitem [{\citenamefont {Cao}\ \emph {et~al.}(2021{\natexlab{b}})\citenamefont
  {Cao}, \citenamefont {Park}, \citenamefont {Watanabe}, \citenamefont
  {Taniguchi},\ and\ \citenamefont {Jarillo-Herrero}}]{cao2021pauli}%
  \BibitemOpen
  \bibfield  {author} {\bibinfo {author} {\bibfnamefont {Y.}~\bibnamefont
  {Cao}}, \bibinfo {author} {\bibfnamefont {J.~M.}\ \bibnamefont {Park}},
  \bibinfo {author} {\bibfnamefont {K.}~\bibnamefont {Watanabe}}, \bibinfo
  {author} {\bibfnamefont {T.}~\bibnamefont {Taniguchi}}, \ and\ \bibinfo
  {author} {\bibfnamefont {P.}~\bibnamefont {Jarillo-Herrero}},\ }\href@noop {}
  {\bibfield  {journal} {\bibinfo  {journal} {Nature}\ }\textbf {\bibinfo
  {volume} {595}},\ \bibinfo {pages} {526} (\bibinfo {year}
  {2021}{\natexlab{b}})}\BibitemShut {NoStop}%
\bibitem [{\citenamefont {Oh}\ \emph {et~al.}(2021)\citenamefont {Oh},
  \citenamefont {Nuckolls}, \citenamefont {Wong}, \citenamefont {Lee},
  \citenamefont {Liu}, \citenamefont {Watanabe}, \citenamefont {Taniguchi},\
  and\ \citenamefont {Yazdani}}]{oh2021evidence}%
  \BibitemOpen
  \bibfield  {author} {\bibinfo {author} {\bibfnamefont {M.}~\bibnamefont
  {Oh}}, \bibinfo {author} {\bibfnamefont {K.~P.}\ \bibnamefont {Nuckolls}},
  \bibinfo {author} {\bibfnamefont {D.}~\bibnamefont {Wong}}, \bibinfo {author}
  {\bibfnamefont {R.~L.}\ \bibnamefont {Lee}}, \bibinfo {author} {\bibfnamefont
  {X.}~\bibnamefont {Liu}}, \bibinfo {author} {\bibfnamefont {K.}~\bibnamefont
  {Watanabe}}, \bibinfo {author} {\bibfnamefont {T.}~\bibnamefont {Taniguchi}},
  \ and\ \bibinfo {author} {\bibfnamefont {A.}~\bibnamefont {Yazdani}},\
  }\href@noop {} {\bibfield  {journal} {\bibinfo  {journal} {Nature}\ }\textbf
  {\bibinfo {volume} {600}},\ \bibinfo {pages} {240} (\bibinfo {year}
  {2021})}\BibitemShut {NoStop}%
\bibitem [{\citenamefont {Kim}\ \emph {et~al.}(2022)\citenamefont {Kim},
  \citenamefont {Choi}, \citenamefont {Lewandowski}, \citenamefont {Thomson},
  \citenamefont {Zhang}, \citenamefont {Polski}, \citenamefont {Watanabe},
  \citenamefont {Taniguchi}, \citenamefont {Alicea},\ and\ \citenamefont
  {Nadj-Perge}}]{kim2022evidence}%
  \BibitemOpen
  \bibfield  {author} {\bibinfo {author} {\bibfnamefont {H.}~\bibnamefont
  {Kim}}, \bibinfo {author} {\bibfnamefont {Y.}~\bibnamefont {Choi}}, \bibinfo
  {author} {\bibfnamefont {C.}~\bibnamefont {Lewandowski}}, \bibinfo {author}
  {\bibfnamefont {A.}~\bibnamefont {Thomson}}, \bibinfo {author} {\bibfnamefont
  {Y.}~\bibnamefont {Zhang}}, \bibinfo {author} {\bibfnamefont
  {R.}~\bibnamefont {Polski}}, \bibinfo {author} {\bibfnamefont
  {K.}~\bibnamefont {Watanabe}}, \bibinfo {author} {\bibfnamefont
  {T.}~\bibnamefont {Taniguchi}}, \bibinfo {author} {\bibfnamefont
  {J.}~\bibnamefont {Alicea}}, \ and\ \bibinfo {author} {\bibfnamefont
  {S.}~\bibnamefont {Nadj-Perge}},\ }\href@noop {} {\bibfield  {journal}
  {\bibinfo  {journal} {Nature}\ }\textbf {\bibinfo {volume} {606}},\ \bibinfo
  {pages} {494} (\bibinfo {year} {2022})}\BibitemShut {NoStop}%
\bibitem [{\citenamefont {Liu}\ \emph {et~al.}(2022)\citenamefont {Liu},
  \citenamefont {Zhang}, \citenamefont {Watanabe}, \citenamefont {Taniguchi},\
  and\ \citenamefont {Li}}]{liu2022isospin}%
  \BibitemOpen
  \bibfield  {author} {\bibinfo {author} {\bibfnamefont {X.}~\bibnamefont
  {Liu}}, \bibinfo {author} {\bibfnamefont {N.~J.}\ \bibnamefont {Zhang}},
  \bibinfo {author} {\bibfnamefont {K.}~\bibnamefont {Watanabe}}, \bibinfo
  {author} {\bibfnamefont {T.}~\bibnamefont {Taniguchi}}, \ and\ \bibinfo
  {author} {\bibfnamefont {J.}~\bibnamefont {Li}},\ }\href@noop {} {\bibfield
  {journal} {\bibinfo  {journal} {Nature Physics}\ }\textbf {\bibinfo {volume}
  {18}},\ \bibinfo {pages} {522} (\bibinfo {year} {2022})}\BibitemShut
  {NoStop}%
\bibitem [{\citenamefont {Zhang}\ \emph {et~al.}(2022)\citenamefont {Zhang},
  \citenamefont {Polski}, \citenamefont {Lewandowski}, \citenamefont {Thomson},
  \citenamefont {Peng}, \citenamefont {Choi}, \citenamefont {Kim},
  \citenamefont {Watanabe}, \citenamefont {Taniguchi}, \citenamefont {Alicea}
  \emph {et~al.}}]{zhang2022promotion}%
  \BibitemOpen
  \bibfield  {author} {\bibinfo {author} {\bibfnamefont {Y.}~\bibnamefont
  {Zhang}}, \bibinfo {author} {\bibfnamefont {R.}~\bibnamefont {Polski}},
  \bibinfo {author} {\bibfnamefont {C.}~\bibnamefont {Lewandowski}}, \bibinfo
  {author} {\bibfnamefont {A.}~\bibnamefont {Thomson}}, \bibinfo {author}
  {\bibfnamefont {Y.}~\bibnamefont {Peng}}, \bibinfo {author} {\bibfnamefont
  {Y.}~\bibnamefont {Choi}}, \bibinfo {author} {\bibfnamefont {H.}~\bibnamefont
  {Kim}}, \bibinfo {author} {\bibfnamefont {K.}~\bibnamefont {Watanabe}},
  \bibinfo {author} {\bibfnamefont {T.}~\bibnamefont {Taniguchi}}, \bibinfo
  {author} {\bibfnamefont {J.}~\bibnamefont {Alicea}},  \emph {et~al.},\
  }\href@noop {} {\bibfield  {journal} {\bibinfo  {journal} {Science}\ }\textbf
  {\bibinfo {volume} {377}},\ \bibinfo {pages} {1538} (\bibinfo {year}
  {2022})}\BibitemShut {NoStop}%
\bibitem [{\citenamefont {Park}\ \emph {et~al.}(2022)\citenamefont {Park},
  \citenamefont {Cao}, \citenamefont {Xia}, \citenamefont {Sun}, \citenamefont
  {Watanabe}, \citenamefont {Taniguchi},\ and\ \citenamefont
  {Jarillo-Herrero}}]{park2022robust}%
  \BibitemOpen
  \bibfield  {author} {\bibinfo {author} {\bibfnamefont {J.~M.}\ \bibnamefont
  {Park}}, \bibinfo {author} {\bibfnamefont {Y.}~\bibnamefont {Cao}}, \bibinfo
  {author} {\bibfnamefont {L.-Q.}\ \bibnamefont {Xia}}, \bibinfo {author}
  {\bibfnamefont {S.}~\bibnamefont {Sun}}, \bibinfo {author} {\bibfnamefont
  {K.}~\bibnamefont {Watanabe}}, \bibinfo {author} {\bibfnamefont
  {T.}~\bibnamefont {Taniguchi}}, \ and\ \bibinfo {author} {\bibfnamefont
  {P.}~\bibnamefont {Jarillo-Herrero}},\ }\href@noop {} {\bibfield  {journal}
  {\bibinfo  {journal} {Nature Materials}\ }\textbf {\bibinfo {volume} {21}},\
  \bibinfo {pages} {877} (\bibinfo {year} {2022})}\BibitemShut {NoStop}%
\bibitem [{\citenamefont {Su}\ \emph {et~al.}(2023)\citenamefont {Su},
  \citenamefont {Kuiri}, \citenamefont {Watanabe}, \citenamefont {Taniguchi},\
  and\ \citenamefont {Folk}}]{su2023superconductivity}%
  \BibitemOpen
  \bibfield  {author} {\bibinfo {author} {\bibfnamefont {R.}~\bibnamefont
  {Su}}, \bibinfo {author} {\bibfnamefont {M.}~\bibnamefont {Kuiri}}, \bibinfo
  {author} {\bibfnamefont {K.}~\bibnamefont {Watanabe}}, \bibinfo {author}
  {\bibfnamefont {T.}~\bibnamefont {Taniguchi}}, \ and\ \bibinfo {author}
  {\bibfnamefont {J.}~\bibnamefont {Folk}},\ }\href@noop {} {\bibfield
  {journal} {\bibinfo  {journal} {Nature Materials}\ }\textbf {\bibinfo
  {volume} {22}},\ \bibinfo {pages} {1332} (\bibinfo {year}
  {2023})}\BibitemShut {NoStop}%
\bibitem [{\citenamefont {Zhou}\ \emph {et~al.}(2022)\citenamefont {Zhou},
  \citenamefont {Holleis}, \citenamefont {Saito}, \citenamefont {Cohen},
  \citenamefont {Huynh}, \citenamefont {Patterson}, \citenamefont {Yang},
  \citenamefont {Taniguchi}, \citenamefont {Watanabe},\ and\ \citenamefont
  {Young}}]{zhou2022isospin}%
  \BibitemOpen
  \bibfield  {author} {\bibinfo {author} {\bibfnamefont {H.}~\bibnamefont
  {Zhou}}, \bibinfo {author} {\bibfnamefont {L.}~\bibnamefont {Holleis}},
  \bibinfo {author} {\bibfnamefont {Y.}~\bibnamefont {Saito}}, \bibinfo
  {author} {\bibfnamefont {L.}~\bibnamefont {Cohen}}, \bibinfo {author}
  {\bibfnamefont {W.}~\bibnamefont {Huynh}}, \bibinfo {author} {\bibfnamefont
  {C.~L.}\ \bibnamefont {Patterson}}, \bibinfo {author} {\bibfnamefont
  {F.}~\bibnamefont {Yang}}, \bibinfo {author} {\bibfnamefont {T.}~\bibnamefont
  {Taniguchi}}, \bibinfo {author} {\bibfnamefont {K.}~\bibnamefont {Watanabe}},
  \ and\ \bibinfo {author} {\bibfnamefont {A.~F.}\ \bibnamefont {Young}},\
  }\href@noop {} {\bibfield  {journal} {\bibinfo  {journal} {Science}\ }\textbf
  {\bibinfo {volume} {375}},\ \bibinfo {pages} {774} (\bibinfo {year}
  {2022})}\BibitemShut {NoStop}%
\bibitem [{\citenamefont {Zhang}\ \emph {et~al.}(2023)\citenamefont {Zhang},
  \citenamefont {Polski}, \citenamefont {Thomson}, \citenamefont
  {Lantagne-Hurtubise}, \citenamefont {Lewandowski}, \citenamefont {Zhou},
  \citenamefont {Watanabe}, \citenamefont {Taniguchi}, \citenamefont {Alicea},\
  and\ \citenamefont {Nadj-Perge}}]{zhang2023enhanced}%
  \BibitemOpen
  \bibfield  {author} {\bibinfo {author} {\bibfnamefont {Y.}~\bibnamefont
  {Zhang}}, \bibinfo {author} {\bibfnamefont {R.}~\bibnamefont {Polski}},
  \bibinfo {author} {\bibfnamefont {A.}~\bibnamefont {Thomson}}, \bibinfo
  {author} {\bibfnamefont {{\'E}.}~\bibnamefont {Lantagne-Hurtubise}}, \bibinfo
  {author} {\bibfnamefont {C.}~\bibnamefont {Lewandowski}}, \bibinfo {author}
  {\bibfnamefont {H.}~\bibnamefont {Zhou}}, \bibinfo {author} {\bibfnamefont
  {K.}~\bibnamefont {Watanabe}}, \bibinfo {author} {\bibfnamefont
  {T.}~\bibnamefont {Taniguchi}}, \bibinfo {author} {\bibfnamefont
  {J.}~\bibnamefont {Alicea}}, \ and\ \bibinfo {author} {\bibfnamefont
  {S.}~\bibnamefont {Nadj-Perge}},\ }\href@noop {} {\bibfield  {journal}
  {\bibinfo  {journal} {Nature}\ }\textbf {\bibinfo {volume} {613}},\ \bibinfo
  {pages} {268} (\bibinfo {year} {2023})}\BibitemShut {NoStop}%
\bibitem [{\citenamefont {Holleis}\ \emph {et~al.}(2023)\citenamefont
  {Holleis}, \citenamefont {Patterson}, \citenamefont {Zhang}, \citenamefont
  {Vituri}, \citenamefont {Yoo}, \citenamefont {Zhou}, \citenamefont
  {Taniguchi}, \citenamefont {Watanabe}, \citenamefont {Berg}, \citenamefont
  {Nadj-Perge} \emph {et~al.}}]{holleis2023nematicity}%
  \BibitemOpen
  \bibfield  {author} {\bibinfo {author} {\bibfnamefont {L.}~\bibnamefont
  {Holleis}}, \bibinfo {author} {\bibfnamefont {C.~L.}\ \bibnamefont
  {Patterson}}, \bibinfo {author} {\bibfnamefont {Y.}~\bibnamefont {Zhang}},
  \bibinfo {author} {\bibfnamefont {Y.}~\bibnamefont {Vituri}}, \bibinfo
  {author} {\bibfnamefont {H.~M.}\ \bibnamefont {Yoo}}, \bibinfo {author}
  {\bibfnamefont {H.}~\bibnamefont {Zhou}}, \bibinfo {author} {\bibfnamefont
  {T.}~\bibnamefont {Taniguchi}}, \bibinfo {author} {\bibfnamefont
  {K.}~\bibnamefont {Watanabe}}, \bibinfo {author} {\bibfnamefont
  {E.}~\bibnamefont {Berg}}, \bibinfo {author} {\bibfnamefont {S.}~\bibnamefont
  {Nadj-Perge}},  \emph {et~al.},\ }\href@noop {} {\bibfield  {journal}
  {\bibinfo  {journal} {arXiv preprint arXiv:2303.00742}\ } (\bibinfo {year}
  {2023})}\BibitemShut {NoStop}%
\bibitem [{\citenamefont {Li}\ \emph {et~al.}(2024)\citenamefont {Li},
  \citenamefont {Xu}, \citenamefont {Li}, \citenamefont {Li}, \citenamefont
  {Li}, \citenamefont {Watanabe}, \citenamefont {Taniguchi}, \citenamefont
  {Tong}, \citenamefont {Shen}, \citenamefont {Lu} \emph
  {et~al.}}]{li2024tunable}%
  \BibitemOpen
  \bibfield  {author} {\bibinfo {author} {\bibfnamefont {C.}~\bibnamefont
  {Li}}, \bibinfo {author} {\bibfnamefont {F.}~\bibnamefont {Xu}}, \bibinfo
  {author} {\bibfnamefont {B.}~\bibnamefont {Li}}, \bibinfo {author}
  {\bibfnamefont {J.}~\bibnamefont {Li}}, \bibinfo {author} {\bibfnamefont
  {G.}~\bibnamefont {Li}}, \bibinfo {author} {\bibfnamefont {K.}~\bibnamefont
  {Watanabe}}, \bibinfo {author} {\bibfnamefont {T.}~\bibnamefont {Taniguchi}},
  \bibinfo {author} {\bibfnamefont {B.}~\bibnamefont {Tong}}, \bibinfo {author}
  {\bibfnamefont {J.}~\bibnamefont {Shen}}, \bibinfo {author} {\bibfnamefont
  {L.}~\bibnamefont {Lu}},  \emph {et~al.},\ }\href@noop {} {\bibfield
  {journal} {\bibinfo  {journal} {Nature}\ }\textbf {\bibinfo {volume} {631}},\
  \bibinfo {pages} {300} (\bibinfo {year} {2024})}\BibitemShut {NoStop}%
\bibitem [{\citenamefont {Zhou}\ \emph {et~al.}(2021)\citenamefont {Zhou},
  \citenamefont {Xie}, \citenamefont {Taniguchi}, \citenamefont {Watanabe},\
  and\ \citenamefont {Young}}]{zhou2021superconductivity}%
  \BibitemOpen
  \bibfield  {author} {\bibinfo {author} {\bibfnamefont {H.}~\bibnamefont
  {Zhou}}, \bibinfo {author} {\bibfnamefont {T.}~\bibnamefont {Xie}}, \bibinfo
  {author} {\bibfnamefont {T.}~\bibnamefont {Taniguchi}}, \bibinfo {author}
  {\bibfnamefont {K.}~\bibnamefont {Watanabe}}, \ and\ \bibinfo {author}
  {\bibfnamefont {A.~F.}\ \bibnamefont {Young}},\ }\href@noop {} {\bibfield
  {journal} {\bibinfo  {journal} {Nature}\ }\textbf {\bibinfo {volume} {598}},\
  \bibinfo {pages} {434} (\bibinfo {year} {2021})}\BibitemShut {NoStop}%
\bibitem [{\citenamefont {Han}\ \emph {et~al.}(2024)\citenamefont {Han},
  \citenamefont {Lu}, \citenamefont {Yao}, \citenamefont {Shi}, \citenamefont
  {Yang}, \citenamefont {Seo}, \citenamefont {Ye}, \citenamefont {Wu},
  \citenamefont {Zhou}, \citenamefont {Liu} \emph
  {et~al.}}]{han2024signatures}%
  \BibitemOpen
  \bibfield  {author} {\bibinfo {author} {\bibfnamefont {T.}~\bibnamefont
  {Han}}, \bibinfo {author} {\bibfnamefont {Z.}~\bibnamefont {Lu}}, \bibinfo
  {author} {\bibfnamefont {Y.}~\bibnamefont {Yao}}, \bibinfo {author}
  {\bibfnamefont {L.}~\bibnamefont {Shi}}, \bibinfo {author} {\bibfnamefont
  {J.}~\bibnamefont {Yang}}, \bibinfo {author} {\bibfnamefont {J.}~\bibnamefont
  {Seo}}, \bibinfo {author} {\bibfnamefont {S.}~\bibnamefont {Ye}}, \bibinfo
  {author} {\bibfnamefont {Z.}~\bibnamefont {Wu}}, \bibinfo {author}
  {\bibfnamefont {M.}~\bibnamefont {Zhou}}, \bibinfo {author} {\bibfnamefont
  {H.}~\bibnamefont {Liu}},  \emph {et~al.},\ }\href@noop {} {\bibfield
  {journal} {\bibinfo  {journal} {arXiv preprint arXiv:2408.15233}\ } (\bibinfo
  {year} {2024})}\BibitemShut {NoStop}%
\bibitem [{\citenamefont {Choi}\ \emph {et~al.}(2024)\citenamefont {Choi},
  \citenamefont {Choi}, \citenamefont {Valentini}, \citenamefont {Patterson},
  \citenamefont {Holleis}, \citenamefont {Sheekey}, \citenamefont {Stoyanov},
  \citenamefont {Cheng}, \citenamefont {Taniguchi}, \citenamefont {Watanabe}
  \emph {et~al.}}]{choi2024electric}%
  \BibitemOpen
  \bibfield  {author} {\bibinfo {author} {\bibfnamefont {Y.}~\bibnamefont
  {Choi}}, \bibinfo {author} {\bibfnamefont {Y.}~\bibnamefont {Choi}}, \bibinfo
  {author} {\bibfnamefont {M.}~\bibnamefont {Valentini}}, \bibinfo {author}
  {\bibfnamefont {C.~L.}\ \bibnamefont {Patterson}}, \bibinfo {author}
  {\bibfnamefont {L.~F.}\ \bibnamefont {Holleis}}, \bibinfo {author}
  {\bibfnamefont {O.~I.}\ \bibnamefont {Sheekey}}, \bibinfo {author}
  {\bibfnamefont {H.}~\bibnamefont {Stoyanov}}, \bibinfo {author}
  {\bibfnamefont {X.}~\bibnamefont {Cheng}}, \bibinfo {author} {\bibfnamefont
  {T.}~\bibnamefont {Taniguchi}}, \bibinfo {author} {\bibfnamefont
  {K.}~\bibnamefont {Watanabe}},  \emph {et~al.},\ }\href@noop {} {\bibfield
  {journal} {\bibinfo  {journal} {arXiv preprint arXiv:2408.12584}\ } (\bibinfo
  {year} {2024})}\BibitemShut {NoStop}%
\bibitem [{\citenamefont {Peltonen}\ \emph {et~al.}(2018)\citenamefont
  {Peltonen}, \citenamefont {Ojaj{\"a}rvi},\ and\ \citenamefont
  {Heikkil{\"a}}}]{peltonen2018mean}%
  \BibitemOpen
  \bibfield  {author} {\bibinfo {author} {\bibfnamefont {T.~J.}\ \bibnamefont
  {Peltonen}}, \bibinfo {author} {\bibfnamefont {R.}~\bibnamefont
  {Ojaj{\"a}rvi}}, \ and\ \bibinfo {author} {\bibfnamefont {T.~T.}\
  \bibnamefont {Heikkil{\"a}}},\ }\href@noop {} {\bibfield  {journal} {\bibinfo
   {journal} {Physical Review B}\ }\textbf {\bibinfo {volume} {98}},\ \bibinfo
  {pages} {220504} (\bibinfo {year} {2018})}\BibitemShut {NoStop}%
\bibitem [{\citenamefont {Wu}\ \emph {et~al.}(2019)\citenamefont {Wu},
  \citenamefont {Hwang},\ and\ \citenamefont {Das~Sarma}}]{wu2019phonon}%
  \BibitemOpen
  \bibfield  {author} {\bibinfo {author} {\bibfnamefont {F.}~\bibnamefont
  {Wu}}, \bibinfo {author} {\bibfnamefont {E.}~\bibnamefont {Hwang}}, \ and\
  \bibinfo {author} {\bibfnamefont {S.}~\bibnamefont {Das~Sarma}},\ }\href@noop
  {} {\bibfield  {journal} {\bibinfo  {journal} {Physical Review B}\ }\textbf
  {\bibinfo {volume} {99}},\ \bibinfo {pages} {165112} (\bibinfo {year}
  {2019})}\BibitemShut {NoStop}%
\bibitem [{\citenamefont {Lian}\ \emph {et~al.}(2019)\citenamefont {Lian},
  \citenamefont {Wang},\ and\ \citenamefont {Bernevig}}]{lian2019twisted}%
  \BibitemOpen
  \bibfield  {author} {\bibinfo {author} {\bibfnamefont {B.}~\bibnamefont
  {Lian}}, \bibinfo {author} {\bibfnamefont {Z.}~\bibnamefont {Wang}}, \ and\
  \bibinfo {author} {\bibfnamefont {B.~A.}\ \bibnamefont {Bernevig}},\
  }\href@noop {} {\bibfield  {journal} {\bibinfo  {journal} {Physical review
  letters}\ }\textbf {\bibinfo {volume} {122}},\ \bibinfo {pages} {257002}
  (\bibinfo {year} {2019})}\BibitemShut {NoStop}%
\bibitem [{\citenamefont {Sarma}\ and\ \citenamefont
  {Wu}(2020)}]{sarma2020electron}%
  \BibitemOpen
  \bibfield  {author} {\bibinfo {author} {\bibfnamefont {S.~D.}\ \bibnamefont
  {Sarma}}\ and\ \bibinfo {author} {\bibfnamefont {F.}~\bibnamefont {Wu}},\
  }\href@noop {} {\bibfield  {journal} {\bibinfo  {journal} {Annals of
  Physics}\ }\textbf {\bibinfo {volume} {417}},\ \bibinfo {pages} {168193}
  (\bibinfo {year} {2020})}\BibitemShut {NoStop}%
\bibitem [{\citenamefont {Shavit}\ \emph {et~al.}(2021)\citenamefont {Shavit},
  \citenamefont {Berg}, \citenamefont {Stern},\ and\ \citenamefont
  {Oreg}}]{shavit2021theory}%
  \BibitemOpen
  \bibfield  {author} {\bibinfo {author} {\bibfnamefont {G.}~\bibnamefont
  {Shavit}}, \bibinfo {author} {\bibfnamefont {E.}~\bibnamefont {Berg}},
  \bibinfo {author} {\bibfnamefont {A.}~\bibnamefont {Stern}}, \ and\ \bibinfo
  {author} {\bibfnamefont {Y.}~\bibnamefont {Oreg}},\ }\href@noop {} {\bibfield
   {journal} {\bibinfo  {journal} {Physical review letters}\ }\textbf {\bibinfo
  {volume} {127}},\ \bibinfo {pages} {247703} (\bibinfo {year}
  {2021})}\BibitemShut {NoStop}%
\bibitem [{\citenamefont {Vi{\~n}as~Bostr{\"o}m}\ \emph
  {et~al.}(2024)\citenamefont {Vi{\~n}as~Bostr{\"o}m}, \citenamefont {Fischer},
  \citenamefont {Profe}, \citenamefont {Zhang}, \citenamefont {Kennes},\ and\
  \citenamefont {Rubio}}]{vinas2024phonon}%
  \BibitemOpen
  \bibfield  {author} {\bibinfo {author} {\bibfnamefont {E.}~\bibnamefont
  {Vi{\~n}as~Bostr{\"o}m}}, \bibinfo {author} {\bibfnamefont {A.}~\bibnamefont
  {Fischer}}, \bibinfo {author} {\bibfnamefont {J.~B.}\ \bibnamefont {Profe}},
  \bibinfo {author} {\bibfnamefont {J.}~\bibnamefont {Zhang}}, \bibinfo
  {author} {\bibfnamefont {D.~M.}\ \bibnamefont {Kennes}}, \ and\ \bibinfo
  {author} {\bibfnamefont {A.}~\bibnamefont {Rubio}},\ }\href@noop {}
  {\bibfield  {journal} {\bibinfo  {journal} {npj Computational Materials}\
  }\textbf {\bibinfo {volume} {10}},\ \bibinfo {pages} {163} (\bibinfo {year}
  {2024})}\BibitemShut {NoStop}%
\bibitem [{\citenamefont {Sharma}\ \emph {et~al.}(2020)\citenamefont {Sharma},
  \citenamefont {Trushin}, \citenamefont {Sushkov}, \citenamefont {Vignale},\
  and\ \citenamefont {Adam}}]{sharma2020superconductivity}%
  \BibitemOpen
  \bibfield  {author} {\bibinfo {author} {\bibfnamefont {G.}~\bibnamefont
  {Sharma}}, \bibinfo {author} {\bibfnamefont {M.}~\bibnamefont {Trushin}},
  \bibinfo {author} {\bibfnamefont {O.~P.}\ \bibnamefont {Sushkov}}, \bibinfo
  {author} {\bibfnamefont {G.}~\bibnamefont {Vignale}}, \ and\ \bibinfo
  {author} {\bibfnamefont {S.}~\bibnamefont {Adam}},\ }\href@noop {} {\bibfield
   {journal} {\bibinfo  {journal} {Physical Review Research}\ }\textbf
  {\bibinfo {volume} {2}},\ \bibinfo {pages} {022040} (\bibinfo {year}
  {2020})}\BibitemShut {NoStop}%
\bibitem [{\citenamefont {Ghazaryan}\ \emph {et~al.}(2021)\citenamefont
  {Ghazaryan}, \citenamefont {Holder}, \citenamefont {Serbyn},\ and\
  \citenamefont {Berg}}]{ghazaryan2021unconventional}%
  \BibitemOpen
  \bibfield  {author} {\bibinfo {author} {\bibfnamefont {A.}~\bibnamefont
  {Ghazaryan}}, \bibinfo {author} {\bibfnamefont {T.}~\bibnamefont {Holder}},
  \bibinfo {author} {\bibfnamefont {M.}~\bibnamefont {Serbyn}}, \ and\ \bibinfo
  {author} {\bibfnamefont {E.}~\bibnamefont {Berg}},\ }\href@noop {} {\bibfield
   {journal} {\bibinfo  {journal} {Physical review letters}\ }\textbf {\bibinfo
  {volume} {127}},\ \bibinfo {pages} {247001} (\bibinfo {year}
  {2021})}\BibitemShut {NoStop}%
\bibitem [{\citenamefont {You}\ and\ \citenamefont
  {Vishwanath}(2022)}]{you2022kohn}%
  \BibitemOpen
  \bibfield  {author} {\bibinfo {author} {\bibfnamefont {Y.-Z.}\ \bibnamefont
  {You}}\ and\ \bibinfo {author} {\bibfnamefont {A.}~\bibnamefont
  {Vishwanath}},\ }\href@noop {} {\bibfield  {journal} {\bibinfo  {journal}
  {Physical Review B}\ }\textbf {\bibinfo {volume} {105}},\ \bibinfo {pages}
  {134524} (\bibinfo {year} {2022})}\BibitemShut {NoStop}%
\bibitem [{\citenamefont {Li}\ \emph {et~al.}(2023)\citenamefont {Li},
  \citenamefont {Kuang}, \citenamefont {Jimeno-Pozo}, \citenamefont
  {Sainz-Cruz}, \citenamefont {Zhan}, \citenamefont {Yuan},\ and\ \citenamefont
  {Guinea}}]{li2023charge}%
  \BibitemOpen
  \bibfield  {author} {\bibinfo {author} {\bibfnamefont {Z.}~\bibnamefont
  {Li}}, \bibinfo {author} {\bibfnamefont {X.}~\bibnamefont {Kuang}}, \bibinfo
  {author} {\bibfnamefont {A.}~\bibnamefont {Jimeno-Pozo}}, \bibinfo {author}
  {\bibfnamefont {H.}~\bibnamefont {Sainz-Cruz}}, \bibinfo {author}
  {\bibfnamefont {Z.}~\bibnamefont {Zhan}}, \bibinfo {author} {\bibfnamefont
  {S.}~\bibnamefont {Yuan}}, \ and\ \bibinfo {author} {\bibfnamefont
  {F.}~\bibnamefont {Guinea}},\ }\href@noop {} {\bibfield  {journal} {\bibinfo
  {journal} {Physical Review B}\ }\textbf {\bibinfo {volume} {108}},\ \bibinfo
  {pages} {045404} (\bibinfo {year} {2023})}\BibitemShut {NoStop}%
\bibitem [{\citenamefont {Cea}\ \emph {et~al.}(2022)\citenamefont {Cea},
  \citenamefont {Pantale{\'o}n}, \citenamefont {Phong},\ and\ \citenamefont
  {Guinea}}]{cea2022superconductivity}%
  \BibitemOpen
  \bibfield  {author} {\bibinfo {author} {\bibfnamefont {T.}~\bibnamefont
  {Cea}}, \bibinfo {author} {\bibfnamefont {P.~A.}\ \bibnamefont
  {Pantale{\'o}n}}, \bibinfo {author} {\bibfnamefont {V.~T.}\ \bibnamefont
  {Phong}}, \ and\ \bibinfo {author} {\bibfnamefont {F.}~\bibnamefont
  {Guinea}},\ }\href@noop {} {\bibfield  {journal} {\bibinfo  {journal}
  {Physical Review B}\ }\textbf {\bibinfo {volume} {105}},\ \bibinfo {pages}
  {075432} (\bibinfo {year} {2022})}\BibitemShut {NoStop}%
\bibitem [{\citenamefont {Chatterjee}\ \emph {et~al.}(2022)\citenamefont
  {Chatterjee}, \citenamefont {Wang}, \citenamefont {Berg},\ and\ \citenamefont
  {Zaletel}}]{chatterjee2022inter}%
  \BibitemOpen
  \bibfield  {author} {\bibinfo {author} {\bibfnamefont {S.}~\bibnamefont
  {Chatterjee}}, \bibinfo {author} {\bibfnamefont {T.}~\bibnamefont {Wang}},
  \bibinfo {author} {\bibfnamefont {E.}~\bibnamefont {Berg}}, \ and\ \bibinfo
  {author} {\bibfnamefont {M.~P.}\ \bibnamefont {Zaletel}},\ }\href@noop {}
  {\bibfield  {journal} {\bibinfo  {journal} {Nature communications}\ }\textbf
  {\bibinfo {volume} {13}},\ \bibinfo {pages} {6013} (\bibinfo {year}
  {2022})}\BibitemShut {NoStop}%
\bibitem [{\citenamefont {Christos}\ \emph {et~al.}(2023)\citenamefont
  {Christos}, \citenamefont {Sachdev},\ and\ \citenamefont
  {Scheurer}}]{christos2023nodal}%
  \BibitemOpen
  \bibfield  {author} {\bibinfo {author} {\bibfnamefont {M.}~\bibnamefont
  {Christos}}, \bibinfo {author} {\bibfnamefont {S.}~\bibnamefont {Sachdev}}, \
  and\ \bibinfo {author} {\bibfnamefont {M.~S.}\ \bibnamefont {Scheurer}},\
  }\href@noop {} {\bibfield  {journal} {\bibinfo  {journal} {Nature
  Communications}\ }\textbf {\bibinfo {volume} {14}},\ \bibinfo {pages} {7134}
  (\bibinfo {year} {2023})}\BibitemShut {NoStop}%
\bibitem [{\citenamefont {Dong}\ \emph {et~al.}(2023)\citenamefont {Dong},
  \citenamefont {Levitov},\ and\ \citenamefont
  {Chubukov}}]{dong2023superconductivity}%
  \BibitemOpen
  \bibfield  {author} {\bibinfo {author} {\bibfnamefont {Z.}~\bibnamefont
  {Dong}}, \bibinfo {author} {\bibfnamefont {L.}~\bibnamefont {Levitov}}, \
  and\ \bibinfo {author} {\bibfnamefont {A.~V.}\ \bibnamefont {Chubukov}},\
  }\href@noop {} {\bibfield  {journal} {\bibinfo  {journal} {Physical Review
  B}\ }\textbf {\bibinfo {volume} {108}},\ \bibinfo {pages} {134503} (\bibinfo
  {year} {2023})}\BibitemShut {NoStop}%
\bibitem [{\citenamefont {Qin}\ \emph {et~al.}(2023)\citenamefont {Qin},
  \citenamefont {Huang}, \citenamefont {Wolf}, \citenamefont {Wei},
  \citenamefont {Blinov},\ and\ \citenamefont {MacDonald}}]{qin2023functional}%
  \BibitemOpen
  \bibfield  {author} {\bibinfo {author} {\bibfnamefont {W.}~\bibnamefont
  {Qin}}, \bibinfo {author} {\bibfnamefont {C.}~\bibnamefont {Huang}}, \bibinfo
  {author} {\bibfnamefont {T.}~\bibnamefont {Wolf}}, \bibinfo {author}
  {\bibfnamefont {N.}~\bibnamefont {Wei}}, \bibinfo {author} {\bibfnamefont
  {I.}~\bibnamefont {Blinov}}, \ and\ \bibinfo {author} {\bibfnamefont {A.~H.}\
  \bibnamefont {MacDonald}},\ }\href@noop {} {\bibfield  {journal} {\bibinfo
  {journal} {Physical Review Letters}\ }\textbf {\bibinfo {volume} {130}},\
  \bibinfo {pages} {146001} (\bibinfo {year} {2023})}\BibitemShut {NoStop}%
\bibitem [{\citenamefont {Dong}\ \emph {et~al.}(2024)\citenamefont {Dong},
  \citenamefont {Lantagne-Hurtubise},\ and\ \citenamefont
  {Alicea}}]{dong2024superconductivity}%
  \BibitemOpen
  \bibfield  {author} {\bibinfo {author} {\bibfnamefont {Z.}~\bibnamefont
  {Dong}}, \bibinfo {author} {\bibfnamefont {{\'E}.}~\bibnamefont
  {Lantagne-Hurtubise}}, \ and\ \bibinfo {author} {\bibfnamefont
  {J.}~\bibnamefont {Alicea}},\ }\href@noop {} {\bibfield  {journal} {\bibinfo
  {journal} {arXiv preprint arXiv:2406.17036}\ } (\bibinfo {year}
  {2024})}\BibitemShut {NoStop}%
\bibitem [{\citenamefont {Fischer}\ \emph {et~al.}(2024)\citenamefont
  {Fischer}, \citenamefont {Klebl}, \citenamefont {Profe}, \citenamefont
  {Rothstein}, \citenamefont {Waldecker}, \citenamefont {Beschoten},
  \citenamefont {Wehling},\ and\ \citenamefont {Kennes}}]{fischer2024spin}%
  \BibitemOpen
  \bibfield  {author} {\bibinfo {author} {\bibfnamefont {A.}~\bibnamefont
  {Fischer}}, \bibinfo {author} {\bibfnamefont {L.}~\bibnamefont {Klebl}},
  \bibinfo {author} {\bibfnamefont {J.~B.}\ \bibnamefont {Profe}}, \bibinfo
  {author} {\bibfnamefont {A.}~\bibnamefont {Rothstein}}, \bibinfo {author}
  {\bibfnamefont {L.}~\bibnamefont {Waldecker}}, \bibinfo {author}
  {\bibfnamefont {B.}~\bibnamefont {Beschoten}}, \bibinfo {author}
  {\bibfnamefont {T.~O.}\ \bibnamefont {Wehling}}, \ and\ \bibinfo {author}
  {\bibfnamefont {D.~M.}\ \bibnamefont {Kennes}},\ }\href@noop {} {\bibfield
  {journal} {\bibinfo  {journal} {Physical Review Research}\ }\textbf {\bibinfo
  {volume} {6}},\ \bibinfo {pages} {L012003} (\bibinfo {year}
  {2024})}\BibitemShut {NoStop}%
\bibitem [{\citenamefont {Khalaf}\ \emph {et~al.}(2021)\citenamefont {Khalaf},
  \citenamefont {Chatterjee}, \citenamefont {Bultinck}, \citenamefont
  {Zaletel},\ and\ \citenamefont {Vishwanath}}]{khalaf2021charged}%
  \BibitemOpen
  \bibfield  {author} {\bibinfo {author} {\bibfnamefont {E.}~\bibnamefont
  {Khalaf}}, \bibinfo {author} {\bibfnamefont {S.}~\bibnamefont {Chatterjee}},
  \bibinfo {author} {\bibfnamefont {N.}~\bibnamefont {Bultinck}}, \bibinfo
  {author} {\bibfnamefont {M.~P.}\ \bibnamefont {Zaletel}}, \ and\ \bibinfo
  {author} {\bibfnamefont {A.}~\bibnamefont {Vishwanath}},\ }\href@noop {}
  {\bibfield  {journal} {\bibinfo  {journal} {Science advances}\ }\textbf
  {\bibinfo {volume} {7}},\ \bibinfo {pages} {eabf5299} (\bibinfo {year}
  {2021})}\BibitemShut {NoStop}%
\bibitem [{\citenamefont {Cea}\ and\ \citenamefont
  {Guinea}(2021)}]{cea2021coulomb}%
  \BibitemOpen
  \bibfield  {author} {\bibinfo {author} {\bibfnamefont {T.}~\bibnamefont
  {Cea}}\ and\ \bibinfo {author} {\bibfnamefont {F.}~\bibnamefont {Guinea}},\
  }\href@noop {} {\bibfield  {journal} {\bibinfo  {journal} {Proceedings of the
  National Academy of Sciences}\ }\textbf {\bibinfo {volume} {118}},\ \bibinfo
  {pages} {e2107874118} (\bibinfo {year} {2021})}\BibitemShut {NoStop}%
\bibitem [{\citenamefont {Kwan}\ \emph {et~al.}(2022)\citenamefont {Kwan},
  \citenamefont {Wagner}, \citenamefont {Bultinck}, \citenamefont {Simon},\
  and\ \citenamefont {Parameswaran}}]{kwan2022skyrmions}%
  \BibitemOpen
  \bibfield  {author} {\bibinfo {author} {\bibfnamefont {Y.~H.}\ \bibnamefont
  {Kwan}}, \bibinfo {author} {\bibfnamefont {G.}~\bibnamefont {Wagner}},
  \bibinfo {author} {\bibfnamefont {N.}~\bibnamefont {Bultinck}}, \bibinfo
  {author} {\bibfnamefont {S.~H.}\ \bibnamefont {Simon}}, \ and\ \bibinfo
  {author} {\bibfnamefont {S.}~\bibnamefont {Parameswaran}},\ }\href@noop {}
  {\bibfield  {journal} {\bibinfo  {journal} {Physical Review X}\ }\textbf
  {\bibinfo {volume} {12}},\ \bibinfo {pages} {031020} (\bibinfo {year}
  {2022})}\BibitemShut {NoStop}%
\bibitem [{\citenamefont {Kim}\ \emph {et~al.}(2025)\citenamefont {Kim},
  \citenamefont {Timmel}, \citenamefont {Ju},\ and\ \citenamefont
  {Wen}}]{kim2025topological}%
  \BibitemOpen
  \bibfield  {author} {\bibinfo {author} {\bibfnamefont {M.}~\bibnamefont
  {Kim}}, \bibinfo {author} {\bibfnamefont {A.}~\bibnamefont {Timmel}},
  \bibinfo {author} {\bibfnamefont {L.}~\bibnamefont {Ju}}, \ and\ \bibinfo
  {author} {\bibfnamefont {X.-G.}\ \bibnamefont {Wen}},\ }\href@noop {}
  {\bibfield  {journal} {\bibinfo  {journal} {Physical Review B}\ }\textbf
  {\bibinfo {volume} {111}},\ \bibinfo {pages} {014508} (\bibinfo {year}
  {2025})}\BibitemShut {NoStop}%
\bibitem [{\citenamefont {Qi}\ and\ \citenamefont
  {Zhang}(2011)}]{qi2011topological}%
  \BibitemOpen
  \bibfield  {author} {\bibinfo {author} {\bibfnamefont {X.-L.}\ \bibnamefont
  {Qi}}\ and\ \bibinfo {author} {\bibfnamefont {S.-C.}\ \bibnamefont {Zhang}},\
  }\href@noop {} {\bibfield  {journal} {\bibinfo  {journal} {Reviews of modern
  physics}\ }\textbf {\bibinfo {volume} {83}},\ \bibinfo {pages} {1057}
  (\bibinfo {year} {2011})}\BibitemShut {NoStop}%
\bibitem [{\citenamefont {Leijnse}\ and\ \citenamefont
  {Flensberg}(2012)}]{leijnse2012introduction}%
  \BibitemOpen
  \bibfield  {author} {\bibinfo {author} {\bibfnamefont {M.}~\bibnamefont
  {Leijnse}}\ and\ \bibinfo {author} {\bibfnamefont {K.}~\bibnamefont
  {Flensberg}},\ }\href@noop {} {\bibfield  {journal} {\bibinfo  {journal}
  {Semiconductor Science and Technology}\ }\textbf {\bibinfo {volume} {27}},\
  \bibinfo {pages} {124003} (\bibinfo {year} {2012})}\BibitemShut {NoStop}%
\bibitem [{\citenamefont {Sato}\ and\ \citenamefont
  {Ando}(2017)}]{sato2017topological}%
  \BibitemOpen
  \bibfield  {author} {\bibinfo {author} {\bibfnamefont {M.}~\bibnamefont
  {Sato}}\ and\ \bibinfo {author} {\bibfnamefont {Y.}~\bibnamefont {Ando}},\
  }\href@noop {} {\bibfield  {journal} {\bibinfo  {journal} {Reports on
  Progress in Physics}\ }\textbf {\bibinfo {volume} {80}},\ \bibinfo {pages}
  {076501} (\bibinfo {year} {2017})}\BibitemShut {NoStop}%
\bibitem [{\citenamefont {Geier}\ \emph {et~al.}(2024)\citenamefont {Geier},
  \citenamefont {Davydova},\ and\ \citenamefont {Fu}}]{geier2024chiral}%
  \BibitemOpen
  \bibfield  {author} {\bibinfo {author} {\bibfnamefont {M.}~\bibnamefont
  {Geier}}, \bibinfo {author} {\bibfnamefont {M.}~\bibnamefont {Davydova}}, \
  and\ \bibinfo {author} {\bibfnamefont {L.}~\bibnamefont {Fu}},\ }\href@noop
  {} {\bibfield  {journal} {\bibinfo  {journal} {arXiv preprint
  arXiv:2409.13829}\ } (\bibinfo {year} {2024})}\BibitemShut {NoStop}%
\bibitem [{\citenamefont {Shavit}\ and\ \citenamefont
  {Alicea}(2024)}]{shavit2024quantum}%
  \BibitemOpen
  \bibfield  {author} {\bibinfo {author} {\bibfnamefont {G.}~\bibnamefont
  {Shavit}}\ and\ \bibinfo {author} {\bibfnamefont {J.}~\bibnamefont
  {Alicea}},\ }\href@noop {} {\bibfield  {journal} {\bibinfo  {journal} {arXiv
  preprint arXiv:2411.05071}\ } (\bibinfo {year} {2024})}\BibitemShut {NoStop}%
\bibitem [{\citenamefont {Jahin}\ and\ \citenamefont
  {Lin}(2024)}]{jahin2024enhanced}%
  \BibitemOpen
  \bibfield  {author} {\bibinfo {author} {\bibfnamefont {A.}~\bibnamefont
  {Jahin}}\ and\ \bibinfo {author} {\bibfnamefont {S.-Z.}\ \bibnamefont
  {Lin}},\ }\href@noop {} {\bibfield  {journal} {\bibinfo  {journal} {arXiv
  preprint arXiv:2411.09664}\ } (\bibinfo {year} {2024})}\BibitemShut {NoStop}%
\bibitem [{\citenamefont {Chou}\ \emph {et~al.}(2024)\citenamefont {Chou},
  \citenamefont {Zhu},\ and\ \citenamefont {Sarma}}]{chou2024intravalley}%
  \BibitemOpen
  \bibfield  {author} {\bibinfo {author} {\bibfnamefont {Y.-Z.}\ \bibnamefont
  {Chou}}, \bibinfo {author} {\bibfnamefont {J.}~\bibnamefont {Zhu}}, \ and\
  \bibinfo {author} {\bibfnamefont {S.~D.}\ \bibnamefont {Sarma}},\ }\href@noop
  {} {\bibfield  {journal} {\bibinfo  {journal} {arXiv preprint
  arXiv:2409.06701}\ } (\bibinfo {year} {2024})}\BibitemShut {NoStop}%
\bibitem [{\citenamefont {Yang}\ and\ \citenamefont
  {Zhang}(2024)}]{yang2024topological}%
  \BibitemOpen
  \bibfield  {author} {\bibinfo {author} {\bibfnamefont {H.}~\bibnamefont
  {Yang}}\ and\ \bibinfo {author} {\bibfnamefont {Y.-H.}\ \bibnamefont
  {Zhang}},\ }\href@noop {} {\bibfield  {journal} {\bibinfo  {journal} {arXiv
  preprint arXiv:2411.02503}\ } (\bibinfo {year} {2024})}\BibitemShut {NoStop}%
\bibitem [{\citenamefont {Qin}\ and\ \citenamefont {Wu}(2024)}]{qin2024chiral}%
  \BibitemOpen
  \bibfield  {author} {\bibinfo {author} {\bibfnamefont {Q.}~\bibnamefont
  {Qin}}\ and\ \bibinfo {author} {\bibfnamefont {C.}~\bibnamefont {Wu}},\
  }\href {\doibase 10.48550/arXiv.2412.07145} {\bibfield  {journal} {\bibinfo
  {journal} {arXiv preprint arXiv:2412.07145}\ } (\bibinfo {year} {2024}),\
  10.48550/arXiv.2412.07145}\BibitemShut {NoStop}%
\bibitem [{\citenamefont {Parra-Martinez}\ \emph {et~al.}(2025)\citenamefont
  {Parra-Martinez}, \citenamefont {Jimeno-Pozo}, \citenamefont {Phong},
  \citenamefont {Sainz-Cruz}, \citenamefont {Kaplan}, \citenamefont {Emanuel},
  \citenamefont {Oreg}, \citenamefont {Pantaleon}, \citenamefont
  {Silva-Guillen},\ and\ \citenamefont {Guinea}}]{parra2025band}%
  \BibitemOpen
  \bibfield  {author} {\bibinfo {author} {\bibfnamefont {G.}~\bibnamefont
  {Parra-Martinez}}, \bibinfo {author} {\bibfnamefont {A.}~\bibnamefont
  {Jimeno-Pozo}}, \bibinfo {author} {\bibfnamefont {V.~T.}\ \bibnamefont
  {Phong}}, \bibinfo {author} {\bibfnamefont {H.}~\bibnamefont {Sainz-Cruz}},
  \bibinfo {author} {\bibfnamefont {D.}~\bibnamefont {Kaplan}}, \bibinfo
  {author} {\bibfnamefont {P.}~\bibnamefont {Emanuel}}, \bibinfo {author}
  {\bibfnamefont {Y.}~\bibnamefont {Oreg}}, \bibinfo {author} {\bibfnamefont
  {P.~A.}\ \bibnamefont {Pantaleon}}, \bibinfo {author} {\bibfnamefont {J.~A.}\
  \bibnamefont {Silva-Guillen}}, \ and\ \bibinfo {author} {\bibfnamefont
  {F.}~\bibnamefont {Guinea}},\ }\href@noop {} {\bibfield  {journal} {\bibinfo
  {journal} {arXiv preprint arXiv:2502.19474}\ } (\bibinfo {year}
  {2025})}\BibitemShut {NoStop}%
\bibitem [{\citenamefont {Saito}\ \emph {et~al.}(2020)\citenamefont {Saito},
  \citenamefont {Ge}, \citenamefont {Watanabe}, \citenamefont {Taniguchi},\
  and\ \citenamefont {Young}}]{saito2020independent}%
  \BibitemOpen
  \bibfield  {author} {\bibinfo {author} {\bibfnamefont {Y.}~\bibnamefont
  {Saito}}, \bibinfo {author} {\bibfnamefont {J.}~\bibnamefont {Ge}}, \bibinfo
  {author} {\bibfnamefont {K.}~\bibnamefont {Watanabe}}, \bibinfo {author}
  {\bibfnamefont {T.}~\bibnamefont {Taniguchi}}, \ and\ \bibinfo {author}
  {\bibfnamefont {A.~F.}\ \bibnamefont {Young}},\ }\href@noop {} {\bibfield
  {journal} {\bibinfo  {journal} {Nature Physics}\ }\textbf {\bibinfo {volume}
  {16}},\ \bibinfo {pages} {926} (\bibinfo {year} {2020})}\BibitemShut
  {NoStop}%
\bibitem [{\citenamefont {Stepanov}\ \emph {et~al.}(2020)\citenamefont
  {Stepanov}, \citenamefont {Das}, \citenamefont {Lu}, \citenamefont
  {Fahimniya}, \citenamefont {Watanabe}, \citenamefont {Taniguchi},
  \citenamefont {Koppens}, \citenamefont {Lischner}, \citenamefont {Levitov},\
  and\ \citenamefont {Efetov}}]{stepanov2020untying}%
  \BibitemOpen
  \bibfield  {author} {\bibinfo {author} {\bibfnamefont {P.}~\bibnamefont
  {Stepanov}}, \bibinfo {author} {\bibfnamefont {I.}~\bibnamefont {Das}},
  \bibinfo {author} {\bibfnamefont {X.}~\bibnamefont {Lu}}, \bibinfo {author}
  {\bibfnamefont {A.}~\bibnamefont {Fahimniya}}, \bibinfo {author}
  {\bibfnamefont {K.}~\bibnamefont {Watanabe}}, \bibinfo {author}
  {\bibfnamefont {T.}~\bibnamefont {Taniguchi}}, \bibinfo {author}
  {\bibfnamefont {F.~H.}\ \bibnamefont {Koppens}}, \bibinfo {author}
  {\bibfnamefont {J.}~\bibnamefont {Lischner}}, \bibinfo {author}
  {\bibfnamefont {L.}~\bibnamefont {Levitov}}, \ and\ \bibinfo {author}
  {\bibfnamefont {D.~K.}\ \bibnamefont {Efetov}},\ }\href@noop {} {\bibfield
  {journal} {\bibinfo  {journal} {Nature}\ }\textbf {\bibinfo {volume} {583}},\
  \bibinfo {pages} {375} (\bibinfo {year} {2020})}\BibitemShut {NoStop}%
\bibitem [{\citenamefont {Liu}\ \emph {et~al.}(2021)\citenamefont {Liu},
  \citenamefont {Wang}, \citenamefont {Watanabe}, \citenamefont {Taniguchi},
  \citenamefont {Vafek},\ and\ \citenamefont {Li}}]{liu2021tuning}%
  \BibitemOpen
  \bibfield  {author} {\bibinfo {author} {\bibfnamefont {X.}~\bibnamefont
  {Liu}}, \bibinfo {author} {\bibfnamefont {Z.}~\bibnamefont {Wang}}, \bibinfo
  {author} {\bibfnamefont {K.}~\bibnamefont {Watanabe}}, \bibinfo {author}
  {\bibfnamefont {T.}~\bibnamefont {Taniguchi}}, \bibinfo {author}
  {\bibfnamefont {O.}~\bibnamefont {Vafek}}, \ and\ \bibinfo {author}
  {\bibfnamefont {J.}~\bibnamefont {Li}},\ }\href@noop {} {\bibfield  {journal}
  {\bibinfo  {journal} {Science}\ }\textbf {\bibinfo {volume} {371}},\ \bibinfo
  {pages} {1261} (\bibinfo {year} {2021})}\BibitemShut {NoStop}%
\bibitem [{\citenamefont {Barrier}\ \emph {et~al.}(2024)\citenamefont
  {Barrier}, \citenamefont {Peng}, \citenamefont {Xu}, \citenamefont {Fal'ko},
  \citenamefont {Watanabe}, \citenamefont {Tanigushi}, \citenamefont {Geim},
  \citenamefont {Adam},\ and\ \citenamefont {Berdyugin}}]{barrier2024coulomb}%
  \BibitemOpen
  \bibfield  {author} {\bibinfo {author} {\bibfnamefont {J.}~\bibnamefont
  {Barrier}}, \bibinfo {author} {\bibfnamefont {L.}~\bibnamefont {Peng}},
  \bibinfo {author} {\bibfnamefont {S.}~\bibnamefont {Xu}}, \bibinfo {author}
  {\bibfnamefont {V.}~\bibnamefont {Fal'ko}}, \bibinfo {author} {\bibfnamefont
  {K.}~\bibnamefont {Watanabe}}, \bibinfo {author} {\bibfnamefont
  {T.}~\bibnamefont {Tanigushi}}, \bibinfo {author} {\bibfnamefont
  {A.}~\bibnamefont {Geim}}, \bibinfo {author} {\bibfnamefont {S.}~\bibnamefont
  {Adam}}, \ and\ \bibinfo {author} {\bibfnamefont {A.~I.}\ \bibnamefont
  {Berdyugin}},\ }\href@noop {} {\bibfield  {journal} {\bibinfo  {journal}
  {arXiv preprint arXiv:2412.01577}\ } (\bibinfo {year} {2024})}\BibitemShut
  {NoStop}%
\bibitem [{\citenamefont {Koshino}\ and\ \citenamefont
  {McCann}(2009)}]{koshino2009trigonal}%
  \BibitemOpen
  \bibfield  {author} {\bibinfo {author} {\bibfnamefont {M.}~\bibnamefont
  {Koshino}}\ and\ \bibinfo {author} {\bibfnamefont {E.}~\bibnamefont
  {McCann}},\ }\href {\doibase 10.1103/PhysRevB.80.165409} {\bibfield
  {journal} {\bibinfo  {journal} {Physical Review B}\ }\textbf {\bibinfo
  {volume} {80}},\ \bibinfo {pages} {165409} (\bibinfo {year}
  {2009})}\BibitemShut {NoStop}%
\bibitem [{\citenamefont {Zhang}\ \emph {et~al.}(2010)\citenamefont {Zhang},
  \citenamefont {Sahu}, \citenamefont {Min},\ and\ \citenamefont
  {MacDonald}}]{zhang2010band}%
  \BibitemOpen
  \bibfield  {author} {\bibinfo {author} {\bibfnamefont {F.}~\bibnamefont
  {Zhang}}, \bibinfo {author} {\bibfnamefont {B.}~\bibnamefont {Sahu}},
  \bibinfo {author} {\bibfnamefont {H.}~\bibnamefont {Min}}, \ and\ \bibinfo
  {author} {\bibfnamefont {A.~H.}\ \bibnamefont {MacDonald}},\ }\href@noop {}
  {\bibfield  {journal} {\bibinfo  {journal} {Physical Review B—Condensed
  Matter and Materials Physics}\ }\textbf {\bibinfo {volume} {82}},\ \bibinfo
  {pages} {035409} (\bibinfo {year} {2010})}\BibitemShut {NoStop}%
\bibitem [{\citenamefont {Adams}\ and\ \citenamefont
  {Blount}(1959)}]{adams1959energy}%
  \BibitemOpen
  \bibfield  {author} {\bibinfo {author} {\bibfnamefont {E.}~\bibnamefont
  {Adams}}\ and\ \bibinfo {author} {\bibfnamefont {E.}~\bibnamefont {Blount}},\
  }\href@noop {} {\bibfield  {journal} {\bibinfo  {journal} {Journal of Physics
  and Chemistry of Solids}\ }\textbf {\bibinfo {volume} {10}},\ \bibinfo
  {pages} {286} (\bibinfo {year} {1959})}\BibitemShut {NoStop}%
\bibitem [{\citenamefont {Nagaosa}\ \emph {et~al.}(2010)\citenamefont
  {Nagaosa}, \citenamefont {Sinova}, \citenamefont {Onoda}, \citenamefont
  {MacDonald},\ and\ \citenamefont {Ong}}]{nagaosa2010anomalous}%
  \BibitemOpen
  \bibfield  {author} {\bibinfo {author} {\bibfnamefont {N.}~\bibnamefont
  {Nagaosa}}, \bibinfo {author} {\bibfnamefont {J.}~\bibnamefont {Sinova}},
  \bibinfo {author} {\bibfnamefont {S.}~\bibnamefont {Onoda}}, \bibinfo
  {author} {\bibfnamefont {A.~H.}\ \bibnamefont {MacDonald}}, \ and\ \bibinfo
  {author} {\bibfnamefont {N.~P.}\ \bibnamefont {Ong}},\ }\href@noop {}
  {\bibfield  {journal} {\bibinfo  {journal} {Reviews of modern physics}\
  }\textbf {\bibinfo {volume} {82}},\ \bibinfo {pages} {1539} (\bibinfo {year}
  {2010})}\BibitemShut {NoStop}%
\bibitem [{\citenamefont {Price}\ \emph {et~al.}(2014)\citenamefont {Price},
  \citenamefont {Ozawa},\ and\ \citenamefont {Carusotto}}]{price2014quantum}%
  \BibitemOpen
  \bibfield  {author} {\bibinfo {author} {\bibfnamefont {H.~M.}\ \bibnamefont
  {Price}}, \bibinfo {author} {\bibfnamefont {T.}~\bibnamefont {Ozawa}}, \ and\
  \bibinfo {author} {\bibfnamefont {I.}~\bibnamefont {Carusotto}},\ }\href@noop
  {} {\bibfield  {journal} {\bibinfo  {journal} {Physical review letters}\
  }\textbf {\bibinfo {volume} {113}},\ \bibinfo {pages} {190403} (\bibinfo
  {year} {2014})}\BibitemShut {NoStop}%
\bibitem [{\citenamefont {Little}\ and\ \citenamefont
  {Parks}(1962)}]{PhysRevLett.9.9}%
  \BibitemOpen
  \bibfield  {author} {\bibinfo {author} {\bibfnamefont {W.~A.}\ \bibnamefont
  {Little}}\ and\ \bibinfo {author} {\bibfnamefont {R.~D.}\ \bibnamefont
  {Parks}},\ }\href {\doibase 10.1103/PhysRevLett.9.9} {\bibfield  {journal}
  {\bibinfo  {journal} {Phys. Rev. Lett.}\ }\textbf {\bibinfo {volume} {9}},\
  \bibinfo {pages} {9} (\bibinfo {year} {1962})}\BibitemShut {NoStop}%
\bibitem [{\citenamefont {Read}\ and\ \citenamefont
  {Green}(2000)}]{read2000paired}%
  \BibitemOpen
  \bibfield  {author} {\bibinfo {author} {\bibfnamefont {N.}~\bibnamefont
  {Read}}\ and\ \bibinfo {author} {\bibfnamefont {D.}~\bibnamefont {Green}},\
  }\href {\doibase 10.1103/PhysRevB.61.10267} {\bibfield  {journal} {\bibinfo
  {journal} {Physical Review B}\ }\textbf {\bibinfo {volume} {61}},\ \bibinfo
  {pages} {10267} (\bibinfo {year} {2000})}\BibitemShut {NoStop}%
\bibitem [{\citenamefont {Tan}\ and\ \citenamefont
  {Devakul}(2024)}]{tan2024parent}%
  \BibitemOpen
  \bibfield  {author} {\bibinfo {author} {\bibfnamefont {T.}~\bibnamefont
  {Tan}}\ and\ \bibinfo {author} {\bibfnamefont {T.}~\bibnamefont {Devakul}},\
  }\href@noop {} {\bibfield  {journal} {\bibinfo  {journal} {Physical Review
  X}\ }\textbf {\bibinfo {volume} {14}},\ \bibinfo {pages} {041040} (\bibinfo
  {year} {2024})}\BibitemShut {NoStop}%
\bibitem [{\citenamefont {Resta}(2011)}]{resta2011insulating}%
  \BibitemOpen
  \bibfield  {author} {\bibinfo {author} {\bibfnamefont {R.}~\bibnamefont
  {Resta}},\ }\href@noop {} {\bibfield  {journal} {\bibinfo  {journal} {The
  European Physical Journal B}\ }\textbf {\bibinfo {volume} {79}},\ \bibinfo
  {pages} {121} (\bibinfo {year} {2011})}\BibitemShut {NoStop}%
\bibitem [{\citenamefont {Parameswaran}\ \emph {et~al.}(2013)\citenamefont
  {Parameswaran}, \citenamefont {Roy},\ and\ \citenamefont
  {Sondhi}}]{parameswaran2013fractional}%
  \BibitemOpen
  \bibfield  {author} {\bibinfo {author} {\bibfnamefont {S.~A.}\ \bibnamefont
  {Parameswaran}}, \bibinfo {author} {\bibfnamefont {R.}~\bibnamefont {Roy}}, \
  and\ \bibinfo {author} {\bibfnamefont {S.~L.}\ \bibnamefont {Sondhi}},\
  }\href@noop {} {\bibfield  {journal} {\bibinfo  {journal} {Comptes Rendus
  Physique}\ }\textbf {\bibinfo {volume} {14}},\ \bibinfo {pages} {816}
  (\bibinfo {year} {2013})}\BibitemShut {NoStop}%
\bibitem [{\citenamefont {Roy}(2014)}]{roy2014band}%
  \BibitemOpen
  \bibfield  {author} {\bibinfo {author} {\bibfnamefont {R.}~\bibnamefont
  {Roy}},\ }\href@noop {} {\bibfield  {journal} {\bibinfo  {journal} {Physical
  Review B}\ }\textbf {\bibinfo {volume} {90}},\ \bibinfo {pages} {165139}
  (\bibinfo {year} {2014})}\BibitemShut {NoStop}%
\bibitem [{\citenamefont {Jackson}\ \emph {et~al.}(2015)\citenamefont
  {Jackson}, \citenamefont {M{\"o}ller},\ and\ \citenamefont
  {Roy}}]{jackson2015geometric}%
  \BibitemOpen
  \bibfield  {author} {\bibinfo {author} {\bibfnamefont {T.~S.}\ \bibnamefont
  {Jackson}}, \bibinfo {author} {\bibfnamefont {G.}~\bibnamefont {M{\"o}ller}},
  \ and\ \bibinfo {author} {\bibfnamefont {R.}~\bibnamefont {Roy}},\
  }\href@noop {} {\bibfield  {journal} {\bibinfo  {journal} {Nature
  communications}\ }\textbf {\bibinfo {volume} {6}},\ \bibinfo {pages} {8629}
  (\bibinfo {year} {2015})}\BibitemShut {NoStop}%
\bibitem [{\citenamefont {Kohn}\ and\ \citenamefont
  {Luttinger}(1965)}]{kohn1965new}%
  \BibitemOpen
  \bibfield  {author} {\bibinfo {author} {\bibfnamefont {W.}~\bibnamefont
  {Kohn}}\ and\ \bibinfo {author} {\bibfnamefont {J.}~\bibnamefont
  {Luttinger}},\ }\href@noop {} {\bibfield  {journal} {\bibinfo  {journal}
  {Physical Review Letters}\ }\textbf {\bibinfo {volume} {15}},\ \bibinfo
  {pages} {524} (\bibinfo {year} {1965})}\BibitemShut {NoStop}%
\bibitem [{\citenamefont {Raghu}\ \emph {et~al.}(2010)\citenamefont {Raghu},
  \citenamefont {Kivelson},\ and\ \citenamefont
  {Scalapino}}]{raghu2010superconductivity}%
  \BibitemOpen
  \bibfield  {author} {\bibinfo {author} {\bibfnamefont {S.}~\bibnamefont
  {Raghu}}, \bibinfo {author} {\bibfnamefont {S.}~\bibnamefont {Kivelson}}, \
  and\ \bibinfo {author} {\bibfnamefont {D.}~\bibnamefont {Scalapino}},\
  }\href@noop {} {\bibfield  {journal} {\bibinfo  {journal} {Physical Review
  B—Condensed Matter and Materials Physics}\ }\textbf {\bibinfo {volume}
  {81}},\ \bibinfo {pages} {224505} (\bibinfo {year} {2010})}\BibitemShut
  {NoStop}%
\bibitem [{\citenamefont {Arovas}\ \emph {et~al.}(2022)\citenamefont {Arovas},
  \citenamefont {Berg}, \citenamefont {Kivelson},\ and\ \citenamefont
  {Raghu}}]{arovas2022hubbard}%
  \BibitemOpen
  \bibfield  {author} {\bibinfo {author} {\bibfnamefont {D.~P.}\ \bibnamefont
  {Arovas}}, \bibinfo {author} {\bibfnamefont {E.}~\bibnamefont {Berg}},
  \bibinfo {author} {\bibfnamefont {S.~A.}\ \bibnamefont {Kivelson}}, \ and\
  \bibinfo {author} {\bibfnamefont {S.}~\bibnamefont {Raghu}},\ }\href@noop {}
  {\bibfield  {journal} {\bibinfo  {journal} {Annual review of condensed matter
  physics}\ }\textbf {\bibinfo {volume} {13}},\ \bibinfo {pages} {239}
  (\bibinfo {year} {2022})}\BibitemShut {NoStop}%
\bibitem [{sup()}]{supp}%
  \BibitemOpen
  \href@noop {} {}\bibinfo {note} {See Supplemental Material at
  URL-will-be-inserted-by-publisher for ....}\BibitemShut {Stop}%
\bibitem [{\citenamefont {Chubukov}(1993)}]{chubukov1993kohn}%
  \BibitemOpen
  \bibfield  {author} {\bibinfo {author} {\bibfnamefont {A.~V.}\ \bibnamefont
  {Chubukov}},\ }\href@noop {} {\bibfield  {journal} {\bibinfo  {journal}
  {Physical Review B}\ }\textbf {\bibinfo {volume} {48}},\ \bibinfo {pages}
  {1097} (\bibinfo {year} {1993})}\BibitemShut {NoStop}%
\bibitem [{\citenamefont {Basko}\ and\ \citenamefont
  {Aleiner}(2008)}]{basko2008interplay}%
  \BibitemOpen
  \bibfield  {author} {\bibinfo {author} {\bibfnamefont {D.}~\bibnamefont
  {Basko}}\ and\ \bibinfo {author} {\bibfnamefont {I.}~\bibnamefont
  {Aleiner}},\ }\href@noop {} {\bibfield  {journal} {\bibinfo  {journal}
  {Physical Review B—Condensed Matter and Materials Physics}\ }\textbf
  {\bibinfo {volume} {77}},\ \bibinfo {pages} {041409} (\bibinfo {year}
  {2008})}\BibitemShut {NoStop}%
\bibitem [{\citenamefont {Wu}\ \emph {et~al.}(2018)\citenamefont {Wu},
  \citenamefont {MacDonald},\ and\ \citenamefont {Martin}}]{wu2018theory}%
  \BibitemOpen
  \bibfield  {author} {\bibinfo {author} {\bibfnamefont {F.}~\bibnamefont
  {Wu}}, \bibinfo {author} {\bibfnamefont {A.~H.}\ \bibnamefont {MacDonald}}, \
  and\ \bibinfo {author} {\bibfnamefont {I.}~\bibnamefont {Martin}},\
  }\href@noop {} {\bibfield  {journal} {\bibinfo  {journal} {Physical review
  letters}\ }\textbf {\bibinfo {volume} {121}},\ \bibinfo {pages} {257001}
  (\bibinfo {year} {2018})}\BibitemShut {NoStop}%
\bibitem [{\citenamefont {Chou}\ \emph {et~al.}(2021)\citenamefont {Chou},
  \citenamefont {Wu}, \citenamefont {Sau},\ and\ \citenamefont
  {Das~Sarma}}]{chou2021acoustic}%
  \BibitemOpen
  \bibfield  {author} {\bibinfo {author} {\bibfnamefont {Y.-Z.}\ \bibnamefont
  {Chou}}, \bibinfo {author} {\bibfnamefont {F.}~\bibnamefont {Wu}}, \bibinfo
  {author} {\bibfnamefont {J.~D.}\ \bibnamefont {Sau}}, \ and\ \bibinfo
  {author} {\bibfnamefont {S.}~\bibnamefont {Das~Sarma}},\ }\href@noop {}
  {\bibfield  {journal} {\bibinfo  {journal} {Physical review letters}\
  }\textbf {\bibinfo {volume} {127}},\ \bibinfo {pages} {187001} (\bibinfo
  {year} {2021})}\BibitemShut {NoStop}%
\bibitem [{\citenamefont {Chou}\ \emph
  {et~al.}(2022{\natexlab{a}})\citenamefont {Chou}, \citenamefont {Wu},
  \citenamefont {Sau},\ and\ \citenamefont {Das~Sarma}}]{chou2022acoustic}%
  \BibitemOpen
  \bibfield  {author} {\bibinfo {author} {\bibfnamefont {Y.-Z.}\ \bibnamefont
  {Chou}}, \bibinfo {author} {\bibfnamefont {F.}~\bibnamefont {Wu}}, \bibinfo
  {author} {\bibfnamefont {J.~D.}\ \bibnamefont {Sau}}, \ and\ \bibinfo
  {author} {\bibfnamefont {S.}~\bibnamefont {Das~Sarma}},\ }\href@noop {}
  {\bibfield  {journal} {\bibinfo  {journal} {Physical Review B}\ }\textbf
  {\bibinfo {volume} {105}},\ \bibinfo {pages} {L100503} (\bibinfo {year}
  {2022}{\natexlab{a}})}\BibitemShut {NoStop}%
\bibitem [{\citenamefont {Chou}\ \emph
  {et~al.}(2022{\natexlab{b}})\citenamefont {Chou}, \citenamefont {Wu},
  \citenamefont {Sau},\ and\ \citenamefont {Das~Sarma}}]{chou2022acoustic2}%
  \BibitemOpen
  \bibfield  {author} {\bibinfo {author} {\bibfnamefont {Y.-Z.}\ \bibnamefont
  {Chou}}, \bibinfo {author} {\bibfnamefont {F.}~\bibnamefont {Wu}}, \bibinfo
  {author} {\bibfnamefont {J.~D.}\ \bibnamefont {Sau}}, \ and\ \bibinfo
  {author} {\bibfnamefont {S.}~\bibnamefont {Das~Sarma}},\ }\href@noop {}
  {\bibfield  {journal} {\bibinfo  {journal} {Physical Review B}\ }\textbf
  {\bibinfo {volume} {106}},\ \bibinfo {pages} {024507} (\bibinfo {year}
  {2022}{\natexlab{b}})}\BibitemShut {NoStop}%
\end{thebibliography}%
\bibliographystyle{apsrev4-1}
\clearpage

\title{Supplemental material: How pairing mechanism dictates topology in valley-polarized superconductors with Berry curvature}
\maketitle 
\widetext
\section{Rotation Invariant Gauge for the Bloch Functions}
Here we shall discuss the smooth rotation invariant gauge that we use in the main text. To begin, let us consider a band with Bloch functions $\ket{u(\bm{k})}$ in an arbitrary. Our choice of gauge is uniquely determined by the following requirements. First, we require the gauge to be smooth. Second, in order to make the gauge rotation invariant, we require that
\begin{equation}
     \ket{u(R_\theta \bm{k})} = e^{-i L' \theta } U_\theta \ket{u(\bm{k})}
\end{equation}
where the matrix $U_\theta$ is the rotation operator in the orbital basis for an angle $\theta$, and $R_\theta \bm{k}$ is the vector $\bm{k}$ rotated by $\theta$. The integer $L'$ is needed to ensure that the gauge is smooth at $\bm{k} = 0$. Explicitly, $e^{i L' \theta}$ is the solution to the eigenvalue equation,
\begin{equation}
     U_\theta \ket{u(0)} = e^{i L' \theta }\ket{u(0)}.
\end{equation}
Different choices of $L'$ add vortices at $\bm{k} = 0$. Third, we require that $\ket{u(0)}$ is be real and positive. Fourth and finally, we require that  $\bra{u(0)}\ket{u(\bm{k})}$ is real for $\bm{k} = (k_x,0)$ with $k_x > 0$.

It is useful to point out two important properties of this gauge. First, for $\bm{k}$ and $\bm{k}'$ on the Fermi-surface, the form factors can be written as 
\begin{equation}
    F(\bm{k},\bm{k}') = \bra{u(\bm{k})}\ket{u(\bm{k}')} = \sum_n |u_n(\bm{k})| e^{-i (L_n-L') \theta}
\end{equation}
where $u_n(\bm{k}) \equiv \bra{n}\ket{u(\bm{k})}$, is the projection of the Bloch function onto the $n^{\text{th}}$ microscopic orbital, and $L_n$ is the angular momentum of the $n^{
\text{th}}$ orbital, which satisfies
\begin{equation}
    \bra{n}\ket{u(R_\theta \bm{k})} = e^{i (L_n-L') \theta }\bra{n}\ket{u(\bm{k})}.
\label{eq:transformationBlProj}\end{equation}
This transformation property is a direct consequence of the fact that we are working in a rotation invariant gauge, and that the microscopic orbitals have well-defined angular momentum. 

Second, since the Bloch functions are smooth functions of momentum in this gauge, $ \bra{n}\ket{u(R_\theta \bm{k})}$, can be expanded in powers of $\bm{k}$. In order to have the transformation property in Eq.~\ref{eq:transformationBlProj}, the first non-trivial term in the expansion must be proportional to $(k_x \pm ik_y)^{|(L_n-L')|}$, where the $\pm$ is for $L_n-L' > 0$ and $L_n-L' < 0$ respectively. 

\section{Methods of many-body calculations}
In this section, we will detail the methods used to find the effective overscreened interaction for the constant Berry curvature band, and R$N$G. 

\begin{figure}[h]
\centering
    \includegraphics[width=.5\linewidth]{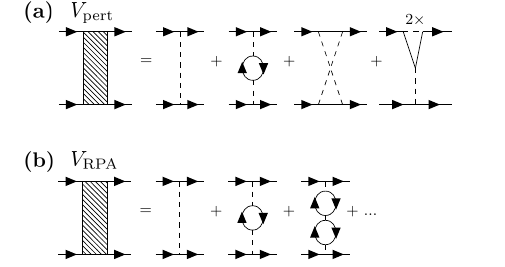}
\caption{Diagrammatic expansion for the effective interaction in the Cooper channel using  \textbf{(a)} perturbation theory and \textbf{(b)} RPA.}
\label{fig:Diagrams}\end{figure}

\subsection{Diagrammatic evaluation }

For the diagrammatic analysis, we are using the Hamiltonian
\begin{equation}
    \begin{split}
         \hat{H} &= \sum_{\bm{q}} \epsilon(\bm{q}) \gamma^\dagger_{\bm{q}}\gamma^{\phantom{\dagger}}_{\bm{q}}  + \hat{H}_{\text{int}}, \\
    \hat{H}_{\text{int}} &= \!\!\!\sum_{\bm{q},\bm{k}_1,\bm{k}_2} \!\!\! \frac{1}{2A}V^{\phantom{\dagger}}_{\bm{q}} \phantom{|} F^{\phantom{\dagger}}_{\bm{k}_1,\bm{k}_1\text{-}\bm{q}} \phantom{|}F^{\phantom{\dagger}}_{\bm{k}_2,\bm{k}_2+\bm{q}} \phantom{|}\gamma^\dagger_{\bm{k}_1}\gamma^{\phantom{\dagger}}_{\bm{k}_1\text{-}\bm{q}} \gamma^\dagger_{\bm{k}_2}\gamma^{\phantom{\dagger}}_{\bm{k}_2+\bm{q}},
    \end{split}
\end{equation}
where $\gamma^\dagger$ are the band projected creation operators, $\epsilon(\bm{k})$ is the band dispersion, and $\hat{H}_{\text{int}}$ is a band-projected density-density interaction.
To calculate the effective overscreened interaction, we use perturbation theory and the random phase approximation (RPA). The corresponding diagrammatic expansions of the effective interaction in the Cooper channel are shown in Fig. ~\ref{fig:Diagrams}(a) and (b) respectively. The effective particle-particle interaction $V_{\bm{k},\bm{k}^\prime}$ obtained from the diagrammatic approach is projected onto the Fermi surface by setting $\abs{\bm{k}} = \abs{\bm{k}^\prime}=k_F$. In order to find the superconducting order parameter $\Delta(\bm{k})$, we use a BCS mean field approach with the linearized gap equation (LGE) 
\begin{align}
    \Delta(\bm{k}) &= \int \frac{d^2\bm{k}^\prime}{(2\pi)^2} \, V_{\bm{k},\bm{k}^\prime} \, \langle \gamma_{\bm{k}^\prime} \gamma_{-\bm{k}^\prime} \rangle \eqret
    &= - \int \frac{\dd^2 \bm{k}^\prime}{(2\pi)^2} \, V_{\bm{k},\bm{k}^\prime} \, \frac{\tanh(\frac{E_{\bm{k}^\prime}}{2 k_\text{B} T_c})}{2E_{\bm{k}^\prime}} \,\Delta(\bm{k}^\prime) \eqret
    &= -\rho_0 \int_{-\omega}^\omega \dd E \ \frac{\tanh(\frac{E}{2 k_\text{B} T_c})}{2E} \, \int_{0}^{2\pi} \frac{\dd \theta_{\bm{k}^\prime}}{2\pi} \, V(\theta_{\bm{k}}-\theta_{\bm{k}^\prime})\, \Delta(\theta_{\bm{k}^\prime})\eqret
    &\approx -\rho_0 \log(\frac{\omega \,\ee^{-\psi(\frac{1}{2})}}{ 2 \pi \,k_\text{B} T_c})\, \int_{0}^{2\pi} \frac{\dd \theta_{\bm{k}^\prime}}{2\pi} \, V(\theta_{\bm{k}}-\theta_{\bm{k}^\prime}) \,\Delta(\theta_{\bm{k}^\prime})
\end{align}
with an energy cutoff $\omega$ ($ k_\text{B} T_c \ll \omega \ll E_f$) of the quasiparticle energy $E_{\bm{k}} = \sqrt{\epsilon(\bm{k})^2 + \abs{\Delta(\bm{k})}^2}\approx \abs{\epsilon(\bm{k})}$ linearized around the critical temperature $T_c$.
$\theta_{\bm{k}}$ and $\theta_{\bm{k^\prime}}$ denote the polar angles of $\bm{k}$ and $\bm{k}^\prime$, respectively, while $\rho_0$ is the density of states at the Fermi level and $\psi(z)$ is the digamma function.
The LGE decouples in terms of the angular momentum channels $l\in \mathbb{Z}$ with the gap function decomposed in terms of angular momentum components $\Delta_l$,
\begin{align}
    \Delta(\theta_{\bm{k}}) = \sum_{l=0}^\infty \Delta_l \,\ee^{- \ii l \theta_{\bm{k}}},
    \label{eq:gap_function}
\end{align}
as 
\begin{align}
    \Delta_l =  -\rho_0 \log(\frac{\omega \,\ee^{-\psi(\frac{1}{2})}}{ 2 \pi \,k_\text{B} T_c})\, \int_{0}^{2\pi} \frac{\dd \theta_{\bm{k}^\prime}}{2\pi} \, V(\theta_{\bm{k}}-\theta_{\bm{k}^\prime}) \,\Delta_l \,  \ee^{\ii l (\theta_{\bm{k^\prime}}-\theta_{\bm{k}})} \equiv -\rho_0 \log(\frac{\omega \,\ee^{-\psi(\frac{1}{2})}}{ 2 \pi \,k_\text{B} T_c})\,\Delta_l \, V_l.
    \label{eq:LGE}
\end{align}
For $V_l < 0 $, where a Fermi instability occurs, the critical temperature $T_c$ for the $l$-channel is then given by
\begin{align}
    k_\text{B} T_c = \frac{\ee^{-\psi(\frac{1}{2})}}{2\pi} \, \omega \, \ee^{-\frac{1}{\rho_0 \, \abs{V_l}}}.
\end{align}
For spinless fermions, only the odd angular momentum components in~\eqref{eq:gap_function} contribute, such that $\Delta_l = 0 $ for $l$ even.
From the Cooper logarithm in~\eqref{eq:LGE}, the largest $T_c$ is determined by the most negative $V_l$, which is identified as the dominant superconducting channel.
In equation~\eqref{eq:LGE}, $V_l$ corresponds to the eigenvalue obtained by diagonalization of the interaction on the Fermi surface through a discrete Fourier transformation as
\begin{align}
  V_{l} &= \frac{1}{(2\pi)^2} \int_{0}^{2\pi} \dd \theta_{\bm{k}} \, \int_{0}^{2\pi} \dd \theta_{\bm{k^\prime}} \,
  \ee^{\ii l (\theta_{\bm{k^\prime}}-\theta_{\bm{k}})} \,  V(\theta_{\bm{k}}-\theta_{\bm{k^\prime}})
  = \frac{1}{2\pi} \int_{0}^{2\pi} \dd \theta \, \ee^{-\ii l \theta} \, V(\theta),
  \label{eq:eff_int}
\end{align}
where $\theta \equiv \theta_{\bm{k}} - \theta_{\bm{k^\prime}}$. The $l$-channel components in~\eqref{eq:eff_int} are analyzed in the following to extract the dominant superconducting channel.

\subsection{Evaluation for the constant Berry curvature band}
First, we consider the diagrammatric approaches for the band of constant Berry curvature and dispersion $\epsilon(\bm{k}) = |\bm{k}|^2/2m$, where the form factor is given by
\begin{equation}
    F(\bm{k_1},\bm{k}_2) =  e^{-\tfrac{\mathcal{B} + \sigma^2}{4}|\bm{k}_1-\bm{k}_2|^2 - i \tfrac{\mathcal{B}}{2} \left( \bm{k}_1\times\bm{k}_2 \right)}.
\label{eq:non-idealFF}\end{equation}
To determine the dominant SC instability, we sample $200$ points equally spaced along the Fermi-surface. 

\subsubsection{Bare interaction}
The bare interaction (corresponding to the first diagram on the RHS of Fig.~\ref{fig:Diagrams}a) is equal to
\begin{align}
  V_{\bm{k},\bm{k^\prime}} = U_0 \, F(\bm{k},\bm{k^\prime})F(-\bm{k},-\bm{k^\prime})
  \label{eq:first_order_Vkk}
\end{align}
with the form factor defined in Eq.~\eqref{eq:non-idealFF}. 
The effective interaction in the $l$-channel is obtained by inserting this into the angular momentum projection~\eqref{eq:eff_int} reading
\begin{align}
  \label{eq:V_first_order}
  V_{l} = \int_{0}^{2\pi}  \frac{\dd\theta}{2\pi} \ \ee^{-\ii l \theta + [\mathcal{B}+\sigma^2] k_F^2 \, (\cos(\theta)-1) +\ii \, \mathcal{B} k_F^2 \sin(\theta)}.
\end{align}
Such an integral of the form
\begin{align}
    \int_{0}^{2\pi}  \frac{\dd\theta}{2\pi} \ \ee^{\ii l \theta} \, u(\theta) \, w(\theta)
\end{align}
with periodic functions $u(\theta)$ with $u(2\pi) = u(0)$ and $w(\theta)$ with $w(2\pi) = w(0)$ can be solved exactly by making use of the discrete convolution theorem
\begin{align}
  \mathcal{F}\{u\cdot w\}(l) = \left[U \star W \right](l).
  \label{eq:discrete_conv_theorem}
\end{align}
Here,
\begin{align}
  U_l = \mathcal{F}\{u\}(l) = \frac{1}{2\pi}\, \int_{0}^{2\pi} \dd \theta \, \ee^{\ii l \theta} \, u(\theta).
\end{align}
with $l\in \mathbb{Z}$ denotes the discrete Fourier transform of $u(\theta)$. Correspondingly, $W_l$ are the Fourier components of the function $w(\theta)$. The discrete convolution of $U_l$ and $W_l$ used in~\eqref{eq:discrete_conv_theorem} is given by
\begin{align}
  \left[U \star W \right](l) = \sum_{k=-\infty}^\infty U_k W_{l-k} = \sum_{k=-\infty}^\infty W_{l-k} V_{k} .
  \label{eq:discrete_convolution}
\end{align}
We identify the Fourier transformation of $\displaystyle u(\theta) = \ee^{[\mathcal{B}+\sigma^2] k_F^2 \, \cos(\theta)}$ as the modified Bessel function of the first kind, 
\begin{align}
I_l([\mathcal{B}+\sigma^2] k_F^2)=\ii^{-l} \, J_{l}(\ii [\mathcal{B}+\sigma^2] k_F^2),
\end{align}
and that of $\displaystyle  v(\theta) = \ee^{\ii \, \mathcal{B} k_F^2 \, \sin(\theta)}$ as the Bessel function of the first kind, $\displaystyle  J_{-l}(\mathcal{B} k_F^2)$. We then employ the multiplication theorem over a field of characteristic zero,
\begin{align}
  \frac{1}{\lambda^\nu} \, J_\nu(\lambda x) = \sum_{n=0}^\infty \frac{1}{n!}\left(\frac{(1-\lambda^2) x}{2}\right)^n J_{\nu+n}(x),
  \label{eq:multiplication_theorem}
\end{align}
to find an expression for the modified Bessel function
\begin{align}
  I_l \left([\mathcal{B}+\sigma^2] \,k_F^2 \right) = \sum_{n=0}^{\infty} \frac{\left([\mathcal{B}+\sigma^2]\, k_F^2\right)^n}{n!}   \, J_{l+n}\left([\mathcal{B}+\sigma^2]\, k_F^2\right)
\end{align}
in terms of Bessel functions $J_n(x)$.
Putting everything together and using $V_l^\text{eff} = \mathcal{F}\{ V(\theta)\} (-l)$, we obtain the result
\begin{align}
  V_{l} &= U_0 \, \ee^{-[\mathcal{B}+\sigma^2] \, k_F^2} \, \sum_{n=0}^\infty \sum_{k=-\infty}^\infty \frac{\left([\mathcal{B}+\sigma^2]\, k_F^2\right)^n}{n!} \, J_{k+n}([\mathcal{B}+\sigma^2] \, k_F^2) \, J_{k+l}(\mathcal{B} \, k_F^2) \eqret
  &= U_0 \, \ee^{-[\mathcal{B}+\sigma^2] \, k_F^2} \, \sum_{n=0}^\infty  \frac{\left([\mathcal{B}+\sigma^2]\, k_F^2\right)^n}{n!} \, J_{n-l}(\sigma^2 k_F^2), \eqret
  &= U_0 \, \ee^{-[\mathcal{B}+\sigma^2] \, k_F^2} \, \left(\sqrt{\frac{2\mathcal{B}+\sigma^2}{\sigma^2}}\right)^l \, I_l\left(\sqrt{2 \mathcal{B}\sigma^2+\sigma^4} \, k_F^2\right)
  \label{eq:Vbare_result}
\end{align}
where we made use of the identity
\begin{align}
  \sum_{\nu=-\infty}^\infty J_\nu(x) J_{\nu+n}(y)= J_n(y-x)
  \label{eq:Bessel_identity}
\end{align}
to obtain the second line.
For the case $\sigma^2=0$,  we use the identity
\begin{align}
    \sum_{\nu=-\infty}^\infty J_\nu(x) J_{\nu+n}(x) = \delta_{n,0}.
    \label{eq:Bessel_identity_delta}
\end{align}
and obtain
\begin{align}
  V_{l} =\begin{cases}
    \displaystyle U_0 \, \ee^{-\mathcal{B} k_F^2} \, \frac{(\mathcal{B} \,k_F^2)^l}{l!} & \quad l\geq0,\\
    0 & \quad l<0.\\
  \end{cases}
  \label{eq:Vbare_analytic}
\end{align}
In this case, we observe, that the effective bare interaction in the negative angular momentum channels is zero, while the sign of the interaction in the positive angular momentum channels inherits the sign of $U_0$.
In the large $l$ limit, the modified Bessel function has the approximate form
\begin{align}
  I_\nu(z) \sim \frac{1}{\sqrt{2\pi \abs{\nu}}} \, \left(\frac{\ee z}{2 \abs{\nu}}\right)^\abs{\nu} \sim \frac{1}{\abs{\nu}!} \left(\frac{z}{2}\right)^{\abs{\nu}}.
\end{align}
This gives rise to the asymptotic large $\l$ form of the result~\eqref{eq:Vbare_result} as
\begin{align}
  V_{l} \sim U_0 \, \ee^{-[\mathcal{B}+\sigma^2] k_F^2}
  \begin{cases}
    \frac{1}{l!}\left((\mathcal{B}+\frac{\sigma^2}{2})\, k_F^2 \right)^l & l>0, \\[1.5ex]
    \frac{1}{\abs{l}!}\left(\frac{\sigma^2}{2} \, k_F^2 \right)^\abs{l} & l<0,
  \end{cases}
\end{align}
which agrees with the result in~\eqref{eq:Vbare_analytic} for $\sigma^2=0$. Notice that, for $\sigma^2 >0$, the magnitude of the interaction $\abs{V_l}$ is larger for positive ($+\abs{l}$) channels than for the corresponding negative ($-\abs{l}$) channels.

\subsubsection{Particle-hole bubble diagram}
The particle-hole bubble diagram (second diagram on the RHS of Fig.~\ref{fig:Diagrams}a) is given by
\begin{align}
  \label{eq:diag_5}
  V_{\text{PHB},\bm{k},\bm{k^\prime}} = - U_0^2 \int \frac{\dd^2 \bm{q}}{(2\pi)^2} \, F(\bm{k},\bm{k^\prime}) \,F(\bm{q}+\bm{k^\prime}-\bm{k},\bm{q}) F(-\bm{k},-\bm{k^\prime}) \,F(\bm{q},\bm{q}+\bm{k^\prime}-\bm{k}) \,G_{\bm{q}} \, G_{\bm{q}+\bm{k^\prime}-\bm{k}}
\end{align}
At zero temperature, the Greens function product entering the diagrams is given by
\begin{align}
 G_{\bm{q}} \, G_{\bm{q}+\bm{k}} = - \frac{n(\epsilon_{\bm{k}+\bm{q}})-n(\epsilon_{\bm{q}})}{\epsilon_{\bm{k}+\bm{q}}-\epsilon_{\bm{q}}},
\end{align}
where $\epsilon_{\bm{k}} = \epsilon(\bm{k})$ is the dispersion and $n(\epsilon_{\bm{k}})$ is the Fermi-Dirac distribution. We consider the interaction at the Fermi surface where $\abs{\bm{k}} = \abs{\bm{k^\prime}} = k_F$ and diagonalize it by performing a Fourier transformation to the angular momentum channels $\l\in\mathbb{Z}$ in Eq.~\eqref{eq:eff_int}. Since in this diagram with the form factors given in~\eqref{eq:non-idealFF}, the integrals over $\theta$ and over $\bm{q}$ factorize, we have
\begin{align}
  V_{\text{PHB},l} = - &\frac{U_0^2}{2\pi} \, \int_{0}^{2\pi} \, \dd \theta \,\chi_0(\abs{\bm{k}-\bm{k^\prime}}) \,  \ee^{-\ii l \theta-[\mathcal{B}+\sigma^2 ]\, k_F^2 (2-2\cos(\theta)) + \ii \mathcal{B} k_F^2 \sin(\theta)}.
  \label{eq:Vph_explicit}
\end{align}
$\chi_0(\bm{k}) = \chi_0(\abs{\bm{k}})$ is the bare polarizability of the 2DEG given by
\begin{align}
  \chi_0(\bm{k}) &= \int \frac{\dd^2 q }{(2\pi)^2} \, G_{\bm{q}}G_{\bm{q}+\bm{k}}  \eqret
  &= -\int \frac{\dd^2 q }{(2\pi)^2} \, n(\epsilon_{\bm{q}}) \left[\frac{1}{\epsilon_{\bm{q}}-\epsilon_{\bm{q}-\bm{k}}}-\frac{1}{\epsilon_{\bm{q}+\bm{k}}-\epsilon_{\bm{q}}}\right] \eqret
  &= \frac{1}{2\pi^2} \int \dd^2 q \ \Theta(k_F-\abs{\bm{q}}) \, \frac{1}{\epsilon_{\bm{q}-\bm{k}}-\epsilon_{\bm{q}}} \eqret
  &= \frac{m}{\pi^2} \int_0^{q_\text{max}} \dd q \, q \int_0^{2\pi} \dd \phi  \, \frac{1}{\abs{\bm{k}}^2 - 2 q \abs{\bm{k}} \cos(\phi )},
  \label{eq:polarizability_2DEG}
\end{align}
with $\phi$ being the angle between $\bm{q}$ and $\bm{k}$. The upper bound of the integral over $q$ is determined by $q_\text{max} = \min(k_F,\frac{\abs{\bm{k}}}{2})$, which follows from the identity
\begin{align}
  \int_0^{2\pi} \dd \theta \, \frac{1}{1-x^2 \cos(\theta)^2} = 0 \quad \text{for } x>1.
\end{align}
Through the substitution $x=\frac{2q}{\abs{\bm{k}}}$, we obtain the integral
\begin{align*}
    \chi_0(\bm{k}) &= \frac{m}{(2\pi)^2} \int_0^{\min(1,\frac{2q}{\abs{\bm{k}}})} \dd x \, x \int_0^{2\pi} \dd \phi \, \frac{1}{1-x \, \cos(\phi)},
\end{align*}
which is solved through the Weierstrass substitution $u=\sqrt{\frac{1+x}{1-x}}\,\tan(\frac{\phi}{2})$ as
\begin{align*}
  \int_0^{2\pi} \dd \phi \, \frac{1}{1-x \, \cos(\phi)} &= 2 \int_0^{\pi} \dd \phi \ \frac{1}{1-x \, \cos(\phi)} = \frac{4}{\sqrt{1-x^2}} \int_0^{\infty} \dd u \, \frac{1}{1+u^2}
   = \frac{2\pi}{\sqrt{1-x^2}}.
\end{align*}
Inserting this into~\eqref{eq:polarizability_2DEG}, we obtain the final result
\begin{align}
  \chi_0(\bm{k}) = \frac{m}{2\pi} \int_0^{\min(1,\frac{2q}{\abs{\bm{k}}})} \dd x \, \frac{x}{\sqrt{1-x^2}} =
    \begin{cases}
      \frac{m}{2\pi} &\text{if } \abs{\bm{k}} \leq 2 k_F \\
      \frac{m}{2\pi}\bigg(1-\sqrt{1-\left(\frac{2 k_F}{\abs{\bm{k}}}\right)^2}\bigg)
      &\text{if } \abs{\bm{k}} > 2 k_F \\
    \end{cases}.
\end{align}
Since $\abs{\bm{k}-\bm{k^\prime}} \leq 2 k_F$, the bare polarizability in the integral~\eqref{eq:Vph_explicit} is constant and we can evaluate the integral by employing the discrete convolution theorem in~\eqref{eq:discrete_convolution}.
The integral~\eqref{eq:Vph_explicit} for the particle-hole bubble diagram is then proportional to that of the bare interaction in~\eqref{eq:V_first_order} upon the substitution $[\mathcal{B}+\sigma^2]\rightarrow 2\, [\mathcal{B}+\sigma^2]$ and up to a prefactor, which is constant in $\theta$. This results in
\begin{align}
  V_{\text{PHB},l} &= -\frac{m U_0^2}{2\pi}  \, \ee^{-2[\mathcal{B}+\sigma^2] \, k_F^2} \, \sum_{n=0}^\infty \sum_{k=-\infty}^\infty \frac{\left(2[\mathcal{B}+\sigma^2]\, k_F^2\right)^n}{n!} J_{k+n}(2[\mathcal{B}+\sigma^2] \, k_F^2) \, J_{k+l}(\mathcal{B} \, k_F^2) \\
  &= -\frac{m U_0^2}{2\pi}  \, \ee^{-2[\mathcal{B}+\sigma^2] \, k_F^2} \, \sum_{n=0}^\infty  \frac{\left(2 [\mathcal{B}+\sigma^2]\, k_F^2\right)^n}{n!} J_{n-l}((\mathcal{B}+2\sigma^2 )\, k_F^2)
\end{align}
by making use of the identity~\eqref{eq:Bessel_identity}. In a representation with a single Bessel function, the result is
\begin{align}
 V_{\text{PHB},l} = -\frac{m U_0^2}{2\pi} \, &\ee^{-2[\mathcal{B}+\sigma^2] \, k_F^2} \, \left(\sqrt{\frac{3\mathcal{B}+2\sigma^2}{\mathcal{B}+2\sigma^2}}\right)^l  I_l\left(\sqrt{3\mathcal{B}^2+8\mathcal{B}\sigma^2+4\sigma^4} \, k_F^2\right).
\end{align}
In the large $l$ limit, the asymptotic form of the result reads
\begin{align}
  V_{l} &\sim -\frac{m U_0^2}{2\pi}  \, \ee^{-2[\mathcal{B}+\sigma^2]\, k_F^2}   \begin{cases}
    \frac{1}{l!}\left((\frac{3 \mathcal{B}}{2}+\sigma^2)\, k_F^2 \right)^l & l>0, \\[1.5ex]
    \frac{1}{\abs{l}!}\left((\frac{\mathcal{B}}{2}+\sigma^2)\, k_F^2 \right)^\abs{l} & l<0.
  \end{cases}
\end{align}
Note that all channels in the particle-hole bubble diagram give an attractive effective interaction. Its magnitude is larger for the $+\abs{l}$ than for the $-\abs{l}$ contribution. There is attraction is nonzero for any $\sigma^2$ including the ideal limit $\sigma^2=0$.

\subsubsection{Cross diagram}
The cross diagram (third diagram on the RHS of Fig~\ref{fig:Diagrams}a) does not have a closed analytic form. But it can be reduced down to a single polar integral, which can be easily evaluated numerically. 

The contribution from the cross diagram is
\begin{align}
  V_{\text{Cr}, \bm{k},\bm{k^\prime}} = U_0^2 \int \frac{\dd^2 \bm{q}}{(2\pi)^2} \,  F(\bm{q}+\bm{k}+\bm{k^\prime},\bm{k^\prime}) \,   F(-\bm{k},\bm{q}) F(\bm{k},\bm{q}+ \bm{k}+\bm{k^\prime}) \,  F(\bm{q},-\bm{k^\prime}) \, G_{\bm{q}} \, G_{\bm{q}+\bm{k}+\bm{k^\prime}}.
\end{align}
By reparameterizing the internal momenta as in~\cite{chubukov1993kohn} with $\bm{q} \pm \frac{\bm{p}}{2}$ and $\bm{p} = \bm{k} + \bm{k}^\prime$, we are able to evaluate the integral over $\abs{\bm{q}}$ explicitly. In this parametrization, the particle-particle interaction is given by
\begin{align}
  V_{\text{Cr},\bm{k},\bm{k^\prime}} &= U_0^2\, \int \frac{\dd^2 \bm{q}}{(2\pi)^2} \,  F(\bm{q}+\frac{\bm{p}}{2},\bm{k^\prime}) \,   F(-\bm{k},\bm{q}-\frac{\bm{p}}{2})
 F(\bm{k},\bm{q}+\frac{\bm{p}}{2}) \,  F(\bm{q}-\frac{\bm{p}}{2},-\bm{k^\prime}) \, G_{\bm{q}+\frac{\bm{p}}{2}} \,
  G_{\bm{q}-\frac{\bm{p}}{2}} \eqret
  &= U_0^2 \, \ee^{-[\mathcal{B}+\sigma^2]\, k_F^2+\frac{[\mathcal{B}+\sigma^2]}{4}\,\abs*{\bm{k}+\bm{k}^\prime}^2 - \ii \mathcal{B} \, \bm{k} \cross \bm{k}^\prime}  \int \frac{\dd^2 \bm{q}}{(2\pi)^2} \, \ee^{-[\mathcal{B}+\sigma^2] \abs{\bm{q}}^2} \,  G_{\bm{q}+\frac{\bm{p}}{2}} \, G_{\bm{q}-\frac{\bm{p}}{2}}
  \label{eq:cr_vertex}
\end{align}
with the form factors from~\eqref{eq:non-idealFF}. The Greens function product can be further evaluated to
\begin{align}
  G_{\bm{q}+\frac{\bm{p}}{2}} \, G_{\bm{q}-\frac{\bm{p}}{2}} &=  -\frac{n(\epsilon_{\bm{q}+\frac{\bm{p}}{2}})-n(\epsilon_{\bm{q}-
  \frac{\bm{p}}{2}})}{\epsilon_{\bm{q}+\frac{\bm{p}}{2}}-\epsilon_{\bm{q}-\frac{\bm{p}}{2}}} \eqret
  &= - \frac{m}{\bm{q}\cdot \bm{p}} \,\left( n(\epsilon_{\bm{q}+\frac{\bm{p}}{2}})-n(\epsilon_{\bm{q}-\frac{\bm{p}}{2}}) \right)\eqret
  &= - \frac{m}{l \cos(\phi) \, \abs{\bm{k}+ \bm{k}^\prime}} \begin{cases}
    1 & \text{if }\abs{\bm{q}+\frac{\bm{p}}{2}}<k_F \ \& \ \abs{\bm{q}-\frac{\bm{p}}{2}}>k_F \\
    -1 & \text{if } \abs{\bm{q}-\frac{\bm{p}}{2}}<k_F  \ \& \ \abs{\bm{q}+\frac{\bm{p}}{2}}>k_F
  \end{cases}
  \label{eq:cr_Greens_product}
\end{align}
with $\phi$ defining the angle between $\bm{q}$ and $\bm{p}$. Solving the constraints in~\eqref{eq:cr_Greens_product} for $\abs{\bm{q}}$, we obtain
\begin{align}
  q_{\pm} = \pm \frac{\abs{\bm{p}} \cos(\phi)}{2} + \frac{\abs{\bm{p}}}{2} \sqrt{\left(\big(\frac{2k_F}{\abs{\bm{p}}}\big)^2-1\right)^2+(\cos(\phi))^2}.
  \label{eq:q_boundaries}
\end{align}
We parametrize $\bm{q}$ by half polar coordinates with $\cos(\phi)\geq 0$ and $q\in \mathbb{R}$, where $q = \abs{\bm{q}} \, \text{sign}(\cos(\phi))$. The integral over $\bm{q}$ in~\eqref{eq:cr_vertex} can then be evaluated as
\begin{align}
  &\int \frac{\dd^2 \bm{q}}{(2\pi)^2} \, \ee^{-[\mathcal{B}+\sigma^2] \abs{\bm{q}}^2} \,  G_{\bm{q}+\frac{\bm{p}}{2}} \, G_{\bm{q}-\frac{\bm{p}}{2}} = \frac{2 m}{(2\pi)^2\abs{\bm{k}+\bm{k}^\prime}}\, \int_{-\frac{\pi}{2}}^{\frac{\pi}{2}} \frac{\dd \phi}{\cos{\phi}} \,
  \int_{q_-}^{q_+}\dd q\,\ee^{-[\mathcal{B}+\sigma^2] q^2}
\end{align}
with
\begin{align}
  \int_{q_-}^{q_+} \dd q\,\ee^{-[\mathcal{B}+\sigma^2] q^2} = &\sqrt{\frac{\pi}{4[\mathcal{B}+\sigma^2]\,k_F^2}} \, \left(\erf(\sqrt{\mathcal{B}+\sigma^2}\, q_+) -\erf(\sqrt{\mathcal{B}+\sigma^2}\, q_-) \right).
\end{align}
We are interested in the diagrammatic contribution at the Fermi surface for $\abs{\bm{k}} = \abs{\bm{k}^\prime} = k_F$, which is diagonalized by the discrete Fourier transformation to angular momenta in~\eqref{eq:eff_int}. With the parametrization
\begin{align}
  &\abs{\bm{p}} = \abs{\bm{k}+\bm{k}^\prime}= 2 k_F\abs{\cos(\frac{\theta}{2})}, \\
  &\bm{k} \cross \bm{k}^\prime = -k_F^2 \sin(\theta), \\
  &q_{\pm} = \pm k_F \abs{\cos(\frac{\theta}{2})} \cos(\phi)   + k_F\sqrt{\left(\sin(\frac{\theta}{2})\right)^2 +\left(\cos(\frac{\theta}{2})\right)^2 (\cos(\phi))^2},
\end{align}
where $\theta$ is the angle between $\bm{k}$ and $\bm{k}^\prime$, we obtain the angular momentum projection of the cross diagram contribution
\begin{align}
  V_{\text{Cr},l} = \frac{2m\, U_0^2}{(2\pi)^3} \,
  &\int_{-\pi}^{\pi} \dd \theta \, \frac{ \ee^{-\ii l \theta} \,\ee^{[\mathcal{B}+\sigma^2]k_F^2 \,(\cos(\frac{\theta}{2})^2-1) + \ii \mathcal{B} \, k_F^2 \sin(\theta)} }{2 \abs{\cos(\frac{\theta}{2})}}\eqret  \cdot &\int_{-\frac{\pi}{2}}^{\frac{\pi}{2}} \frac{\dd \phi}{\cos(\phi)} \sqrt{\frac{\pi}{4[\mathcal{B}+\sigma^2]\,k_F^2}}  \left(\erf(\sqrt{\mathcal{B}+\sigma^2}\, q_+) - \erf(\sqrt{\mathcal{B}+\sigma^2}\, q_-) \right).
  \label{eq:Vcr_numerics}
\end{align}
We employ Eq.~\eqref{eq:Vcr_numerics} to numerically compute the cross diagram contribution of our diagrammatic analysis with perturbation theory.

\subsubsection{Vertex correction diagrams}
The vertex correction diagrams (fourth diagram on the RHS of Eq.~\ref{fig:Diagrams}a) also does not have a closed form. But, again, we can reduce it down to a single polar integral, which can be computed numerically. 

The contribution from the vertex correction diagrams is 
\begin{align}
  V_{\text{Ve}, \bm{k},\bm{k^\prime}} = U_0^2 \int \frac{\dd^2 \bm{q}}{(2\pi)^2} \, F(\bm{k},\bm{q}-\bm{k^\prime}+\bm{k}) \,F(\bm{q},\bm{k^\prime})
  F(\bm{q}-\bm{k^\prime}+\bm{k},\bm{q}) \,F(-\bm{k},-\bm{k^\prime}) \,G_{\bm{q}} \, G_{\bm{q}-\bm{k^\prime}+\bm{k}}
  \label{eq:vertex1}
\end{align}
and
\begin{align}
  V_{\text{Ve},\bm{k},\bm{k^\prime}} = U_0^2 \int \frac{\dd^2 \bm{q}}{(2\pi)^2} \, F(-\bm{k},\bm{q}+\bm{k^\prime}-\bm{k}) \,F(\bm{q},-\bm{k^\prime})
  F(\bm{k},\bm{k^\prime}) \,F(\bm{q}+\bm{k^\prime}-\bm{k},\bm{q}) \,G_{\bm{q}} \, G_{\bm{q}+\bm{k^\prime}-\bm{k}}
  \label{eq:vertex2}
\end{align}
respectively. Using the form factors in~\eqref{eq:non-idealFF}, we notice that both terms in~\eqref{eq:vertex1} and~\eqref{eq:vertex2} are equal. In the following, we focus on~\eqref{eq:vertex1}, which we reparametrize through the internal momenta $\bm{q}\pm \frac{\bm{p}}{2}$ with $\bm{p} = \bm{k}^\prime- \bm{k}$ as
\begin{align}
  V_{\text{Ve},\bm{k},\bm{k^\prime}} &= U_0^2 \int \frac{\dd^2 \bm{q}}{(2\pi)^2} \, F(\bm{k},\bm{q}- \frac{\bm{p}}{2}) \,F(\bm{q}+\frac{\bm{p}}{2},\bm{k^\prime})
 F(\bm{q}- \frac{\bm{p}}{2},\bm{q}+ \frac{\bm{p}}{2}) \,F(-\bm{k},-\bm{k^\prime}) \,G_{\bm{q}+ \frac{\bm{p}}{2}} \, G_{\bm{q}-\frac{\bm{p}}{2}} \eqret
  &= U_0^2 \, \ee^{-\frac{\mathcal{B}+\sigma^2}{2} \left(\abs{\bm{k}^\prime-\bm{k}}^2+\frac{1}{4} \abs{\bm{k}^\prime+\bm{k}}^2\right)} \int \frac{\dd^2 \bm{q}}{(2\pi)^2} \, \ee^{-\frac{\mathcal{B}+\sigma^2}{2} \left(\abs{\bm{q}}^2 - \bm{q}\cdot(\bm{k}+\bm{k}^\prime)\right)- \ii \mathcal{B} \, \bm{q}\cross (\bm{k}^\prime-\bm{k})}   G_{\bm{q}+ \frac{\bm{p}}{2}} \, G_{\bm{q}-\frac{\bm{p}}{2}}.
  \label{eq:ve_vertex}
\end{align}
We define $\phi$ as the angle between $\bm{q}$ and $\bm{p}$ and parametrize $\bm{q}$ by half polar coordinates with $\cos(\phi)\geq 0$ and $q\in \mathbb{R}$, where $q = \abs{\bm{q}} \, \text{sign}(\cos(\phi))$.
The Greens function product is determined by~\eqref{eq:cr_Greens_product} with the integration boundaries for $q$ in~\eqref{eq:q_boundaries} where $\abs{\bm{p}} = \abs{\bm{k}^\prime - \bm{k}}$.
We choose $\theta$ as the angle between $\bm{k}$ and $\bm{k}^\prime$, which determines the terms in~\eqref{eq:ve_vertex} as
\begin{align*}
  &\bm{q} \cdot (\bm{k}+\bm{k}^\prime) = 2 \abs{\bm{q}} k_F \sin(\phi) \cos(\frac{\theta}{2}) \, \text{sign}\left(\sin(\frac{\theta}{2})\right)\eqret
  & \phantom{\bm{q} \cdot (\bm{k}+\bm{k}^\prime) }= 2 q \, k_F  \cos(\frac{\theta}{2})\,\text{sign}\left(\sin(\frac{\theta}{2})\right)\, \sin(\phi)\, \text{sign}(\cos(\phi)), \\
  &\abs{\bm{p}} = \abs{\bm{k}^\prime -\bm{k}  }  =2 k_F \abs{\sin(\frac{\theta}{2})}, \\
  &\abs{\bm{k} + \bm{k}^\prime } = 2k_F \abs{\cos(\frac{\theta}{2})},\\
  & \bm{q}\cross (\bm{k}^\prime-\bm{k}) = -2 q \, k_F \abs{\sin(\frac{\theta}{2})} \,\sin(\phi)\, \text{sign}(\cos(\phi)).
\end{align*}
We define
\begin{align}
  s :=  \, k_F \left(
  \sqrt{\frac{\mathcal{B}+\sigma^2}{2}}\cos(\frac{\theta}{2})\,\text{sign}\left(\sin(\frac{\theta}{2})\right)\, \sin(\phi)+\ii \frac{\sqrt{2}\, \mathcal{B}}{\sqrt{\mathcal{B}+\sigma^2}}\,\abs{\sin(\frac{\theta}{2})} \,\sin(\phi)    \right)
\end{align}
and perform the integral over $\bm{q}$ in~\eqref{eq:ve_vertex} with
\begin{align}
  &\int \frac{\dd^2 \bm{q}}{(2\pi)^2} \, \ee^{-\frac{\mathcal{B}+\sigma^2}{2}
  \left(\abs{\bm{q}}^2 - \bm{q}\cdot(\bm{k}+\bm{k}^\prime)\right)- \ii \mathcal{B} \,\bm{q}\cross (\bm{k}^\prime-\bm{k})}   G_{\bm{q}+ \frac{\bm{p}}{2}} \, G_{\bm{q}-\frac{\bm{p}}{2}} \eqret
  &= \frac{2 m}{(2\pi)^2\abs{\bm{k}^\prime - \bm{k}}}\, \int_{-\frac{\pi}{2}}^{\frac{\pi}{2}} \frac{\dd \phi}{\cos{\phi}} \,
    \int_{q_-}^{q_+}\dd q\,\ee^{-\frac{\mathcal{B}+\sigma^2}{2} (q^2-2\sqrt{\frac{2}{\mathcal{B}+\sigma^2}} \, s \, q)}
\end{align}
and
\begin{align}
    \int_{q_-}^{q_+}\dd q\,\ee^{-\frac{\mathcal{B}+\sigma^2}{2} (q^2-2\sqrt{\frac{2}{\mathcal{B}+\sigma^2}} \, s \, q)} = \sqrt{\frac{\pi}{2[\mathcal{B}+\sigma^2]\, k_F^2}} \, \ee^{s^2}\,\Bigg[\erf\left(\sqrt{\frac{\mathcal{B}+\sigma^2}{2}}\, q_+-s\right) -\erf\left(\sqrt{\frac{\mathcal{B}+\sigma^2}{2}}\, q_--s\right) \Bigg].
\end{align}
The boundaries of the integration are determined by Eq.~\eqref{eq:q_boundaries} and evaluate to
\begin{align}
    q_{\pm} = \pm k_F \abs{\sin(\frac{\theta}{2})} \cos(\phi)   + k_F\sqrt{\left(\cos(\frac{\theta}{2})\right)^2 +\left(\sin(\frac{\theta}{2})\right)^2 (\cos(\phi))^2},
\end{align}
in the chosen parametrization. To consider the vertex correction diagram at the Fermi surface for $\abs{\bm{k}}=\abs{\bm{k}^\prime} = k_F$, which we decompose in terms of angular momentum contributions. The Fourier transformation of the interaction in Eq.~\eqref{eq:eff_int} is given by
\begin{align}
  V_{\text{Ve},l} = \frac{2 m \,U_0^2}{(2\pi)^3}\, &\int_{-\pi}^\pi \dd \theta
  \,\frac{\ee^{-\ii l \theta-\frac{[\mathcal{B}+\sigma^2]k_F^2}{2}\left(4\sin(\frac{\theta}{2})^2+\cos(\frac{\theta}{2})^2\right)}}{2 \abs{\sin(\frac{\theta}{2})}}\eqret
  &\int_{-\frac{\pi}{2}}^{\frac{\pi}{2}} \frac{\dd \phi}{\cos{\phi}} \,
   \sqrt{\frac{\pi}{2[\mathcal{B}+\sigma^2]\, k_F^2}} \, \ee^{s^2}\,\Bigg[\erf\left(\sqrt{\frac{\mathcal{B}+\sigma^2}{2}}\, q_+-s\right) -\erf\left(\sqrt{\frac{\mathcal{B}+\sigma^2}{2}}\, q_--s\right) \Bigg].
  \label{eq:Vve_numerics}
\end{align}
We use~\eqref{eq:Vve_numerics} to numerically find the angular momentum contribution of the vertex correction diagram in the $l$-th channel.

\subsubsection{RPA}
Using our previous analysis of the particle-hole bubble, we have that the RPA overscreened interaction is 
\begin{align}
  \label{eq:V_first_order}
  V_{\text{RPA}, l} = \int_{0}^{2\pi}  \frac{\dd\theta}{2\pi} \ \frac{U_0 \ee^{-\ii l \theta + [\mathcal{B}+\sigma^2] k_F^2 \, (\cos(\theta)-1) +\ii \, \mathcal{B} k_F^2 \sin(\theta)}}{ 1 + \frac{m}{2\pi} U_0 \ee^{[\mathcal{B}+\sigma^2] k_F^2 \, (\cos(\theta)-1) } }.
\end{align}

\subsection{Evaluation for R$4$G}
For $h_{\text{R}N\text{G}}$, defined as in the main text, there is not an analytic form for the electronic form factors. Therefore, we have to evaluate the overscreened interactions numerically. In the RPA treatment, 
\begin{equation}
    V_{\text{RPA},l} = \frac{1}{2\pi} \int_{0}^{2\pi} \dd \theta \, \ee^{-\ii l \theta} \, \frac{\tilde{V}(\theta)}{1+\tilde{V}(\theta)\chi_0(\theta)},
\end{equation}
where $\tilde{V}(\theta)$ is the band projected interaction, \begin{align}
  \tilde{V}(\theta) = V_{\bm{k}-\bm{k}'} \, F(\bm{k},\bm{k^\prime})F(-\bm{k},-\bm{k^\prime})
\end{align}
where $|\bm{k}| = |\bm{k}'| = k_F$, and $\theta$ is the angle between $\bm{k}$ and $\bm{k}'$. $\chi_0(\theta)$ is bare polarizability 
\begin{equation}
     \chi_0(\theta) = \int \frac{\dd^2 \bm{q} }{(2\pi)^2} \, G_{\bm{q}}G_{\bm{q}+[\bm{k}-\bm{k}']}
\label{eq:chiDefRNG}\end{equation}
where $\theta$ is the angle between $\bm{k}$ and $\bm{k}'$, i.e., $|\bm{k}-\bm{k}'| = k_F \sqrt{(2-2\cos(\theta))}$.

It is straightforward to calculate $\tilde{V}(\theta)$, owing to the rotation symmetry of the problem. $\chi_0(\theta)$ must be evaluated numerically. We sample $30$ equally spaced values of $\theta$. and for each value of $\theta$, we compute the integral over $\bm{q}$ in Eq.~\ref{eq:chiDefRNG} on a $3000\times 3000$ grid with $-3k_F \leq q_x,q_y\leq 3k_F$.

\section{Spin$-1/2$ Constant Berry Curvature band }

\begin{figure}[t!]
    \centering
    \includegraphics[width=.5\linewidth]{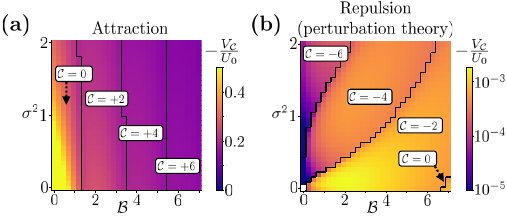}
    \caption{ Strength and BdG Chern number of the dominant interaction channel for spin-1/2 electrons as a function of $\mathcal{B}$ and $\sigma^2$ for \textbf{(a)} short-range attraction and \textbf{(b)} weak short-range repulsion ($\rho U_0 = .1$) treated perturbatively.}
    \label{fig:ICB_PH_S12} 
\end{figure}

In this section, we calculate the dominant superconducting channel for a constant Berry curvature band composed of spin$-1/2$ degrees of freedom. Calculation are done in the same method described in the main text. Results are shown in Fig.~\ref{fig:ICB_PH_S12}. Compared to the spinless case, the particle hole bubbles associated with spin$-1/2$ fermions carry an extra factor of $2$ for the two spin species.

\section{Positive BdG Chern number superconductivity from overscreened interactions.}
\begin{figure*}[t!]
    \centering
    \includegraphics[width=.8\linewidth]{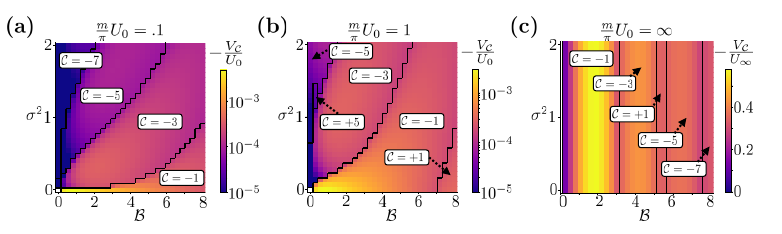}
    \caption{Strength and BdG Chern number of the dominant SC channel for the constant Berry curvature band, as a function of $\mathcal{B}$ and $\sigma^2$. Here we consider repulsive interactions of strength $\frac{m}{\pi} U_0 = .1$, $1$, and $\infty$, and calculate the effective interaction using RPA. Here $U_{\infty} = \frac{2\pi}{m}$.}
    \label{fig:app_ICB_RPA} 
\end{figure*}

In this section, we will present a counter-example to the trend we have established in the main text. Do to this, we will consider the same constant Berry curvature band with a short-ranged repulsive interaction from the main text. This time, we will calculate the overscreened interaction using the random phase approximation (RPA). The results are shown in Fig.~\ref{fig:app_ICB_RPA}. 

For weak interactions ($\frac{m}{\pi}U_0 = .1$) we only find instabilities with $\mathcal{C} < 0$, as in the main text.  However, for $\frac{m}{\pi}U_0 = 1$, and $\infty$, there are regions with $\mathcal{C} > 0$. The small $ \mathcal{C} = + 5$, region that is found for $\frac{m}{\pi}U_0 = 1$, appears to be closely compete with the $\mathcal{C} = -5$, and occurs in a parameter regime where superconductivity is highly suppressed. A $\mathcal{C} = + 1$, region is found for both $\frac{m}{\pi}U_0 = 1$, and $\infty$, at larger values of $\mathcal{B}$. However, the $\mathcal{C} = +1$ regime is a narrow strip of parameter space in the large coupling limit. This can be directly determined from the analytic form of the effective interaction in the $U_0 \rightarrow \infty$,
\begin{equation}
    V_{\mathcal{C}} = \frac{2\pi}{m}J_{\mathcal{C}}(\mathcal{B}),
\end{equation}
which depends only on the Berry curvature $\mathcal{B}$.

As we noted in the main text, $\mathcal{C} = +1$ cooper pairs generically have a non-zero onsite occupancy. However, in the ideal limit is scales as $e^{-\mathcal{B}} \mathcal{B}$, and therefore comes vanishingly when $\mathcal{B}$ becomes large. This allows for $\mathcal{C} = +1$ SC to be energetically competitive with SCs with $\mathcal{C} < 0$. While we do observe positive $\mathcal{C}$ superconductivity in some of these the calculations, the \text{strongest} SC always corresponds to $\mathcal{C} = -1$.

\section{Pairing with an annular Fermi-surfaces}
Let us consider pairing in a system with two Fermi-surfaces, a smaller one at $k_{f-}$ and a larger one at $k_{f+}$. We will assume rotation symmetry throughout. If we assume pairing near the two Fermi-surfaces, we will have two gap functions, $\Delta_{-}$ and $\Delta_{+}$. The self-consitency equations for the two gap functions can be written as 
\begin{equation}\begin{split}
    \Delta_{+}(\bm{k}_{+}) = &\int \frac{d^2\bm{k}'_{+}}{(2\pi)^2}\tilde{V}_{\bm{k}_{+},\bm{k}'_+} \langle \gamma_{\bm{k}'_+} \gamma_{-\bm{k}'_+} \rangle\\ &+ \int \frac{d^2\bm{k}'_{-}}{(2\pi)^2}\tilde{V}_{\bm{k}_{+},\bm{k}'_-} \langle \gamma_{\bm{k}'_-} \gamma_{-\bm{k}'_-} \rangle,\\
    \Delta_{-}(\bm{k}_{-}) = &\int \frac{d^2\bm{k}'_{+}}{(2\pi)^2}\tilde{V}_{\bm{k}_{-},\bm{k}'_+} \langle \gamma_{\bm{k}'_+} \gamma_{-\bm{k}'_+} \rangle\\ &+ \int \frac{d^2\bm{k}'_{-}}{(2\pi)^2}\tilde{V}_{\bm{k}_{-},\bm{k}'_-} \langle \gamma_{\bm{k}'_-} \gamma_{-\bm{k}'_-} \rangle,
\end{split}\end{equation}
where $\gamma$ are the band-projected annihilation operators. Momentum with a $-$ subscript are located near the inner Fermi-surface, and those with a $+$ subscript are located near the outer Fermi-surface. $\tilde{V}$ is the band-projected interaction in the Cooper (particle-particle) channel. For a bare attractive interaction,
\begin{equation}
    \tilde{V}_{\bm{k},\bm{k}'} = V_{\bm{k}-\bm{k}'} F_{\bm{k},\bm{k}'}F_{-\bm{k},-\bm{k}'}.
\end{equation}
 $\tilde{V}_{\bm{k},\bm{k}'}$ can also be an effective interaction obtained, for example, by using RPA to resum particle-hole fluctuations. 

If pairing is only near the Fermi-surface, then we can take $\Delta_{\pm}$ to be independent of $\bm{k}_\pm$, and we can diagonalize the self-consitency equation by considering a gap function with phase winding $l$, $\Delta_{l, \pm}$. For small $\Delta_{l, \pm}$, the resulting linearized gap equation is 
\begin{equation}\begin{split}
    \Delta_{l,+} = &-V_{l,++}\rho_+\log(\frac{W_+}{T}) \Delta_{l,+}\\ &- V_{l,+-}\rho_-\log(\frac{W_-}{T}) \Delta_{l,-},\\
    \Delta_{l,-} = &-V_{l,-+}\rho_+\log(\frac{W_+}{T}) \Delta_{l,+}\\ &- V_{l,--}\rho_-\log(\frac{W_-}{T}) \Delta_{l,-},
\end{split}\end{equation}
where $W_{\pm}$ is a high energy cutoff associated with the inner and outer Fermi-surfaces and $\rho_{\pm}$ is the density of states at the inner and outer Fermi-surfaces.
The interactions, $V_{l,\pm \pm}$are
\begin{equation}\begin{split}
&V_{l,+ +} = \frac{1}{2\pi}\int d\theta_{++}\tilde{V}_{\bm{k}_{+},\bm{k}'_{+}} e^{-il\theta_{++}}\\
&V_{l,+ -} = \frac{1}{2\pi}\int d\theta_{+-}\tilde{V}_{\bm{k}_{+},\bm{k}'_{-}} e^{-il\theta_{+-}}\\
&V_{l,- +} = \frac{1}{2\pi}\int d\theta_{-+}\tilde{V}_{\bm{k}_{-},\bm{k}'_{+}} e^{-il\theta_{-+}}\\
&V_{l,--} = \frac{1}{2\pi}\int d\theta_{--}\tilde{V}_{\bm{k}_{-},\bm{k}'_{-}} e^{-il\theta_{--}}
\end{split}\end{equation}
where $\theta_{\pm \pm}$ is the angle between $\bm{k}_{\pm}$ and $\bm{k}'_{\pm}$ with $|\bm{k}_{\pm}| \approx k_{f\pm}$ and $|\bm{k}'_{\pm}| \approx k_{f\pm}$. Here, the superconducting $T_c$ will depend in a complicated way on $V_{l,\pm\pm]}$ and $W_{\pm}$. However, if $W_+ = W_- \equiv W$, and $\rho_+ = \rho_- \equiv \rho_0$ and the density of states is the same at both Fermi-surface then $T_c = W \exp(-1/V_{l,min} \rho_0)$, where $V_{l,\text{min}}$ is the most negative eigenvalue of the matrix, 
\begin{equation}
    V_l = \begin{bmatrix}V_{l,+ +} & V_{l,- +}\\ V_{l,+ -} & V_{l,- -}
    \end{bmatrix}.
\end{equation}

\section{Results for R$N$G with $N = 2$, $3$, $5$, and $6$}
In this section show results for R$N$G with $N = 2$ (AKA Bernal graphene), $3$, $5$, and $6$. As in the main text, the Hamiltonian is the following $2N\times 2N$ matrix
\begin{equation}
    \begin{split}
    &h_{\text{R$N$G}}(\bm{k})_{2n-1,2n} = h_{\text{R$4$G}}^*(\bm{k})_{2n,2n-1} = v_f (k_x + i k_y)\\
    &h_{\text{R$N$G}}(\bm{k})_{2n,2n+1} = h_{\text{R$4$G}}^*(\bm{k})_{2n,2n-1} = -t_\perp\\
    &h_{\text{R$N$G}}(\bm{k})_{2n-1,2n-1} = h_{\text{R$4$G}}(\bm{k})_{2n,2n} = u_n
\label{eq:NlayerRhombo}\end{split}
\end{equation}
for $1 \leq n \leq N$, where $v_f = 10^6$m/s is the Fermi-velocity, and $t_\perp = .38$eV, and $(u_1,....,u_N) = (-u/2,...,u/2)$, evenly spaced by $u/(N-1)$. 

\begin{figure}[t!]
    \centering
    \includegraphics[width=.55\linewidth]{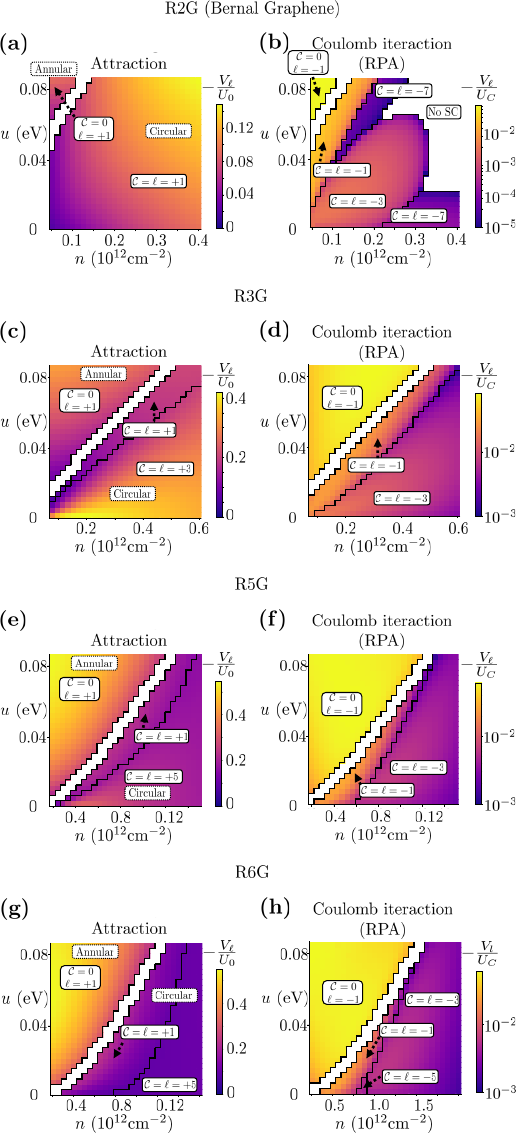}
    \caption{Strength and BdG Chern number of the dominant interaction channel of R$N$G with $N = 2$ (AKA Bernal graphene), $3$, $5$, and $6$. We consider both attractive \textbf{(a,c,e,g)} and overscreened Coulomb interactions (treated using RPA) \textbf{(b,d,f,h)} with $d =20$nm. $\varepsilon = 4$, and $U_C = 1/\varepsilon d$. The regions with annular and circular Fermi-surfaces are separated.}
    \label{fig:RNG_extra} 
\end{figure}

We consider both short-range attraction, and overscreened Coulomb interaction, using the same parameters as the main text. The results are shown in Fig.~\ref{fig:RNG_extra}. For certain parameters, R$2$G (Bernal graphene) did not have an attractive channel with a strength larger than $10^{-5}$ (in relative units). We report "No SC" in these regions.

\end{document}